\documentclass{aa}
\usepackage{graphicx}
\usepackage{txfonts}
\usepackage{natbib}
\usepackage[german,english]{babel}
\begin{document}
\title{EDisCS -- the ESO Distant Cluster Survey\thanks{%
    Based on observations obtained at the ESO Very Large Telescope
    (VLT) as part of the Large Programme 166.A--0162 (the ESO Distant
    Cluster Survey).}}

   \subtitle{Sample Definition and Optical Photometry}

   \author{S.~D.~M.~White\inst{1}, D.~I.~Clowe\inst{2}, L.~Simard\inst{3}, 
   G.~Rudnick\inst{1}, G.~De~Lucia\inst{1}, A.~Arag{\' o}n-Salamanca\inst{4}, 
   R.~Bender\inst{5}, P.~Best\inst{6}, M.~Bremer\inst{7}, S.~Charlot\inst{1},
   J.~Dalcanton\inst{8}, M.~Dantel\inst{9}, V.~Desai\inst{8},
   B.~Fort\inst{10}, C.~Halliday\inst{11}, P.~Jablonka\inst{12},
   G.~Kauffmann\inst{1}, Y.~Mellier\inst{10,9}, B.~Milvang-Jensen\inst{5},
   R.~Pell\'o\inst{13}, B.~Poggianti\inst{14}, S.~Poirier\inst{12},
   H.~Rottgering\inst{15}, R.~Saglia\inst{5}, P.~Schneider\inst{16} \and
   D.~Zaritsky\inst{2}}  


   \offprints{S.~White \email{swhite@mpa-garching.mpg.de}}

   \institute{\selectlanguage{german} Max-Planck-Institut f{\"u}r Astrophysik,
     Karl-Schwarschild-Str. 1, Postfach 1317, D-85741 Garching, Germany
     \selectlanguage{english}
     \and Steward Observatory, University of Arizona, 933 North Cherry Avenue,
     Tucson, AZ 85721  
     \and Herzberg Institute of Astrophysics, National Research Council of
     Canada, Victoria, BC V9E 2E7, Canada  
     \and School of Physics and Astronomy, University of Nottingham,
     University Park, Nottingham NG7 2RD, United Kingdom
     \selectlanguage{german}
     \and Max-Planck Institut f{\"u}r extraterrestrische Physik,
     Giessenbachstrasse, D-85748 Garching, Germany
     \selectlanguage{english}
     \and Institute for Astronomy, Royal Observatory Edinburgh, Blackford Hill,
     Edinburgh EH9 3HJ, UK 
     \and Department of Physics, Bristol University, H. H. Wills Laboratory,
     Tyndall Avenue, Bristol BS8 ITL 
     \and Astronomy Department, University of Washington, Box 351580, Seattle,
     WA 98195  
     \and LERMA, UMR8112, Observatoire de Paris, section de Meudon, 5 Place
     Jules Janssen, F-92195 Meudon Cedex, France
     \and Institut d'Astrophysique de Paris, 98 bis boulevard Arago, 75014
     Paris, France 
     \and Institut f\"ur Astrophysik, Friedrich-Hund-Platz 1, 37077 Goettingen,
     Germany
     \and GEPI, CNRS-UMR8111, Observatoire de Paris, section de Meudon, 5 Place
     Jules Janssen, F-92195 Meudon Cedex, France
     \and Laboratoire d'Astrophysique, UMR 5572, Observatoire Midi-Pyrenees, 14
     Avenue E. Belin, 31400 Toulouse, France
     \and  Osservatorio Astronomico, vicolo dell'Osservatorio 5, 35122
     Padova
     \and Sterrewacht Leiden, P.O. Box 9513, 2300 RA, Leiden, The Netherlands
     \selectlanguage{german} 
     \and Institut f{\"u}r Astrophysik und Extraterrestrische Forschung,
     Universit{\"a}t Bonn, Auf dem H{\"u}gel 71, 53121 Bonn, Germany 
     \selectlanguage{english} 
}

   \authorrunning{S.~D.~M.~White et al.}

   \date{Received ?? ?????? 2005; accepted ?? ????? 2005}

   \abstract{ We present the ESO Distant Cluster Survey (EDisCS) a survey of
     $20$ fields containing distant galaxy clusters with redshifts ranging
     from $0.4$ to almost $1.0$.  Candidate clusters were chosen from among
     the brightest objects identified in the Las Campanas Distant Cluster
     Survey, half with estimated redshift $z_{\rm est}\sim0.5$ and half with
     $z_{\rm est}\sim0.8$.  They were confirmed by identifying red sequences
     in moderately deep two colour data from VLT/FORS2.  For confirmed
     candidates we have assembled deep three-band optical photometry using
     VLT/FORS2, deep near-infrared photometry in one or two bands using
     NTT/SOFI, deep optical spectroscopy using VLT/FORS2, wide field imaging
     in two or three bands using the ESO Wide Field Imager, and HST/ACS mosaic
     images for $10$ of the most distant clusters.  This first paper presents
     our sample and the VLT photometry we have obtained.  We present images,
     colour-magnitude diagrams and richness estimates for our clusters, as
     well as giving redshifts and positions for the brightest cluster members.
     Subsequent papers will present our infrared photometry, spectroscopy, HST
     and wide-field imaging, as well as a wealth of further analysis and
     science results.  Our reduced data become publicly available as these
     papers are accepted.  
      -- Galaxies:
     photometry} 

   \maketitle
%

\section{Introduction}
\label{sec:introduction}

Galaxy clusters were recognised as well defined structures even before the
extragalactic nature of the nebulae was established.  Shortly after Hubble's
monumental discovery they became the systems where the need for unseen ``dark''
matter was first recognised. Sky surveys using the Schmidt telescopes on Mount
Palomar made it possible to compile large catalogues of clusters and these
formed the basis of most cluster studies until X-ray detection of the
intracluster medium became possible for large cluster samples with the Einstein
and ROSAT satellites.  In the near future catalogues based on detection through
gravitational lensing or through inverse Compton scattering of the microwave
background (the Sunyaev-Zel'dovich effect) will provide samples selected
according to cluster mass and thermal energy content, throwing new light on the
relation between galaxies, dark matter, and the intergalactic medium.

Rich galaxy clusters are relatively rare objects and are the most massive
quasi-equilibrium systems in the Universe.  This has led to two different
threads of cluster studies.  One focuses on clusters as the high mass tail of
cosmic structure formation.  Their abundance, clustering and evolution
constrain the amplitude, growth rate and statistics of cosmic density
fluctuations, and hence both the dark matter content of the Universe and the
process responsible for generating all structure within it.  The internal
structure of clusters and their star and gas content constrain the nature of
the dark matter and the cosmic baryon fraction.  Recent years have seen a lot
of work comparing the cosmological constraints provided by cluster studies with
those coming from observations of microwave background fluctuations or of
distant type Ia supernovae.

The second thread studies the galaxies themselves.  Clusters have always played
a major role in galaxy evolution studies.  They provide large samples of
galaxies at the same redshift in a compact and thus easily observable field.
The physical proximity of the galaxies and the dense intracluster medium both
lead to strong environmental effects which may cause many of the striking
differences between ``cluster'' and ``field'' populations. cD galaxies, the
most massive stellar systems known, are found almost exclusively in rich
clusters, which also house the most energetic radio galaxies at the redshifts
where these are most abundant.  Only in clusters is the interaction between
galaxies and intergalactic gas directly observable.  Cluster studies led to the
first direct and unambiguous detection of evolution in the ``normal'' galaxy
population, and clusters continue to provide some of the most compelling
evidence for the wholesale morphological transformation of galaxies.  Finally,
clusters provide a critical test of galaxy formation models based on the now
standard hierarchical paradigm for the growth of structure.

That galaxies in clusters differ systematically from those in the ``field'' was
originally noticed by Hubble and was explored quantitatively in the 1970's by
\citet{O74} and \citet{D80}.  Nearby rich clusters contain more
ellipticals and S0's and fewer star-forming galaxies than poorer systems, a
trend which is particularly marked for concentrated clusters which typically
have high X-ray luminosities and central dominant galaxies. This variation can
be viewed as a relation between morphology and the local space density of
nearby galaxies. Evolution of cluster populations was discovered by
\citet{BO78}; already by $z\sim0.3$ many (but not all) clusters show a
substantial population of blue galaxies not present in significant numbers in
nearby systems. Spectroscopy shows some of these objects to be actively forming
stars, while others have strong Balmer absorption but no emission lines,
indicating the presence of young stars without any current star formation
\citep{DG83}.

Much recent work has concentrated on the nature of these blue cluster members.
The large HST-based study of the ``Morphs'' group
\citep{Detal97,Detal99,Petal99} showed them to have primarily disk-like
morphologies and a very wide range of star formation rates. This study also
found a population of red spiral galaxies and showed S0 galaxies to be less
common in $z\sim 0.5$ than in nearby clusters. In contrast, the abundance of
ellipticals is similar in nearby and distant clusters and their colours and
kinematics suggest purely passive evolution of their stellar populations
\citep{O71,H85,AECC93,BZB96,Ketal98,vDF01}.  These properties can also be
mimicked, however, by models where ellipticals assemble much later
\citep{KC98} and data on the mass growth of central cluster galaxies may even
favour such a model \citep{ABK98}.

A variety of mechanisms have been suggested to explain these trends.  The
relative lack of disk-dominated and star-forming systems was ascribed to
galaxy collisions by \citet{SB51} and to interactions with intracluster gas by
\citet{GG72}. The first process, the transformation of disk galaxies into
spheroidals by tidal shocks, was simulated in some detail by \citet{FS81}, and
was baptised ``galaxy harassment'' in the much later numerical study of
\citet{Metal96}. The second process, ``ram pressure stripping'', was first
simulated by \citet{G76} for ellipticals and has recently been simulated more
realistically by \citet{QMB00} for spirals.  The build-up of massive central
galaxies by repeated accretion of other cluster members (``cannibalism'') was
proposed by \citet{OT75} and \citet{W76a}, while \citet{S76} suggested that
these objects form from the cooling flows which often surround them.  The
formation of normal ellipticals through disk-galaxy mergers was proposed by
\citet{TT72} and has been simulated by many authors beginning with
\citet{FS82} and \citet{NW83}.  \citet{LTC80} suggested that cluster galaxies
may redden because removal of an external  gas supply causes them to run out of
fuel for star formation (``strangulation''), while \citet{DG83} proposed that
starbursts truncate star formation in galaxies that are suddenly overpressured
by the intracluster medium.

Understanding the roles of these various processes requires a comparison of the
observational data with detailed theoretical models for the evolution of
clusters and the galaxies within them.  Early simulations of cluster formation
demonstrated how the regularity of a cluster can be considered a measure of its
dynamical age \citep{W76b} and two decades later the dynamical build-up of cD
galaxies in small clusters could be simulated quite realistically \citep{D98}.
Simulations which can follow the morphology and stellar content of rich cluster
galaxies in the CDM cosmogony have become possible only recently
\citep{Ketal99,Setal01}. A first detailed comparison with a modern cluster
survey was presented by \citet{Detal01}. Many of the environmentally driven
processes discussed above still remain to be incorporated in such work,
however (see \citet{Lanz05} for a first treatment).
 
Since the early 1980's many systematic studies of clusters have been based on
X-ray selected catalogues. For example, the Einstein Medium Sensitivity Survey
\citep{Getal90} provided the base catalogue for the CNOC1 survey of clusters
at $z\sim 0.4$ \citep{Cetal96}, as well as many of the objects in the Morphs
HST survey at similar redshift and the cluster MS1054-03 at $z=0.83$, until
recently the highest redshift cluster studied in any detail
\citep[e.g][]{vDetal99}.  Similarly many cluster catalogues have been built
from ROSAT data \citep{Betal98,Retal95,Mullis03} and these are providing
candidates for ongoing studies of distant clusters, including RXJ0848.9+4452
at $z=1.26$ which has now taken over from MS1054-03 as the most distant
well-studied cluster \citep{Retal99}. Substantial studies of optically
selected samples have been rarer. Recent wide area optical surveys for distant
cluster candidates include those of \citet{Petal96,Petal98}, \citet{dCetal99}
and \citet{Getal01} which cover $6$, $16$, $17$ and $135$ square degrees
respectively. Very recently, \citet{Glad05} have published candidate lists for
the first 10\% of their Red Sequence Cluster Survey which covers $100$ square
degrees. At lower redshifts, the 2dF Galaxy Redshift Survey and the Sloan
Digital Sky Survey provide very large samples of uniformly observed galaxy
clusters which are ideal for studying the systematic properties of cluster
galaxy populations \citep{Gomez03,Bahc03,dePro04}.

Optical selection of distant clusters has a disadvantage with respect to X-ray
selection, in that the incidence of ``false positive'' candidates caused by
line-of-sight superposition of unrelated systems is significantly higher. Thus
most of the $9$ clusters observed by \citet{OPL98} are superpositions of a
number of peaks, and for about half of them the existence of a real cluster
appears dubious.  In contrast none of the $16$ (lower redshift) CNOC1 clusters
has a substantial secondary peak and at most two of the clusters appear
dubious. This superposition problem becomes more serious at higher redshift
because the contrast of clusters with respect to the foreground becomes lower.
It can be reduced but not eliminated by tuning the cluster detection algorithm
to be sensitive to the dense inner regions of clusters and to galaxies with the
colour of elliptical galaxies at the (unknown) cluster redshift.

An optimised filter was used to detect cluster candidates in the Las Campanas
Distant Cluster Survey (LCDCS) on which our own survey is based
\citep{Getal01}.  We chose the LCDCS as the basis for our ESO Distant Cluster
Survey (EDisCS) because it was, at the beginning of our project, by far the
largest uniformly covered area over which one could detect clusters (either at
optical or X-ray wavelengths) to redshifts of $0.9$ or so. Its nominal volume
at $z>0.7$ is larger than the entire local volume within $z=0.1$. We wished to
select samples of rich clusters at $z\sim 0.5$ and $z\sim 0.8$ which were
large and diverse enough to be statistically representative and for which we
could obtain sufficient data to compare the galaxy populations in detail with
those in nearby clusters.  Our principal goal is to understand quantitatively
the evolution of the cluster galaxy population over the last half of the
Hubble time. Optical selection is an advantage here, since similarly selected
nearby cluster samples can be constructed and analysed using the Sloan Digital
Sky Survey.  Our data provide an interesting comparison to other recent
distant cluster surveys which are primarily X-ray selected.

The data we have obtained in the EDisCS programme are of uniform and very high
quality, and they allow us not only to discuss the structure, stellar
populations and kinematics of cluster galaxies out to $z\sim 0.8$, but also to
carry out weak and strong lensing programmes, AGN searches, field studies and a
variety of other programmes which help characterise our clusters and global
galaxy evolution since $z\sim 0.8$. In Sect.~\ref{sec:programme} we set out the
EDisCS programme in more detail and we go through our sample selection and
confirmation procedures.  Our optical photometry, obtained using FORS2 at the
VLT, is presented in Sect.~\ref{sec:photometry}.  We take a first look at our
clusters in Sect.~\ref{sec:sample}, showing images, colour-magnitude diagrams
and morphology plots and giving positions and redshifts for all of them, as
well as discussing each cluster individually.  We give a brief summary in
Sect.~\ref{sec:conclusion}. Our infrared photometry, optical spectroscopy, HST
and wide-field imaging are discussed in later papers of this series, as are a
wide variety of scientific results.

\section{The EDisCS Programme}
\label{sec:programme}

The parent catalogue of our ESO Distant Cluster Survey (EDisCS) is the Las
Campanas Distant Cluster Survey of \citet{Getal01}.  The LCDCS region is a
strip $90$ degrees long and $1.5$ degrees wide centred at $RA=
12.5$ hours and $\delta = -12\degr$. This strip was observed with a
purpose-built drift-scan camera attached to the Las Campanas Observatory $1$m
telescope. A very wide filter ($\sim 4\,500$ to $\sim 7\,700$\AA) maximised the
signal-to-noise of distant clusters against the sky. Each point appears in at
least two drift scans, and the observing technique allows the elimination of
cosmic rays and bad CCD columns, as well as excellent flat-fielding and sky
subtraction.  Resolved objects are detected and replaced with locally drawn
random sky pixels.  Saturated stars and large galaxies are also masked. Any
remaining structure in the image is due to spatial variation in the background
light.  Large-scale structure is removed by subtracting a heavily smoothed
version of the image itself (filter radius $70\farcs$). Clusters are then
detected by convolving with an exponential kernel matched to the expected core
size of galaxy clusters at $z\sim 0.8$ (an exponential scale radius of
$10\farcs$). The limiting surface brightness for a $5\sigma$ detection of a
cluster is roughly $28$ mag/sq.arcsec in I (for ${\rm R}-{\rm
  I}=1$\footnote{Throughout this paper we use Vega magnitudes.}) and just over
$1\,000$ candidate clusters with estimated redshifts in the range $0.3$ to
$1.0$ are detected at this level; about $15$ per cent have estimated redshifts
beyond $0.8$. Redshifts can be estimated almost to $z=0.9$ from the apparent
magnitude of the brightest cluster member.  At higher redshifts even the BCG is
not individually detected, even though the integrated cluster brightness may
still be easily measurable.

The EDisCS programme is a detailed spectroscopic and photometric survey of the
galaxy populations of $10$ among the highest surface brightness clusters in the
LCDCS in each of the ranges $0.45 < z_{\rm est} < 0.55$ and $0.75 < z_{\rm
est} < 0.85$. In the following we will refer to these as the ``mid-$z$'' and
``hi-$z$'' samples respectively.  We aimed for $10$ clusters at each redshift
both because of the substantial cluster-to-cluster variation in galaxy
populations seen at $z\sim 0.4$ (cf the CNOC and MORPHs work) and because of
our wish to obtain statistics for relatively rare cluster populations (cDs,
radio galaxies, other AGN\ldots). Two redshift bins seemed a minimum to see
evolution directly in our own data. Samples of $15$ to $20$ bright candidate
clusters were selected from the LCDCS lists in each redshift range. Each field
was checked carefully to ensure that the detection appeared reliable and free
from observational artifacts. These candidate lists were then followed up in
order of their estimated brightness in a structured observing programme for
which seven phases are now complete.

{\it Phase I: Confirmation Photometry:} We observed our two candidate lists in
LCDCS luminosity order using FORS2 at the VLT.  We first obtained $20$ minute
exposures of 30 fields at I and either at V (mid-$z$ sample) or at R (hi-$z$
sample). We used these to confirm the presence of an apparent cluster with a
brightest member of approximately the magnitude expected.  This was necessary
since in most cases very few galaxies were detected individually in the
original LCDCS drift scans. We also checked for the presence of a possible red
sequence of early-type galaxies at the expected location in the
colour-magnitude diagram.  Candidates failing either of these tests were
rejected and replaced with the next brightest candidate. In practice it was
necessary to observe 13 ``mid-$z$'' LCDCS fields and 17 ``hi-$z$'' fields in
order to obtain $10$ photometrically confirmed clusters in each estimated
redshift range. The statistics of these observations and their implications
for the completeness of the LCDCS are analysed in \citet{Getal02}.

{\it Phase II: Deep Optical Photometry:} We then obtained deep FORS2
photometry at B, V and I for a final set of $10$ photometrically confirmed
mid-$z$ clusters and at V, R and I for a final set of $10$ photometrically
confirmed hi-$z$ clusters. Our total integration times were typically $45$
minutes at the lower redshift and $2$ hours at the higher redshift. Image
quality in all exposures was excellent. The FORS2 field is $6\farcm8\times
6\farcm8$ with a pixel size of $0\farcs20$. After dithering, the field of view
with the maximum depth of exposure was approximately $6.5^{\prime} \times
6.5^{\prime}$.  These are the data presented in more detail in this first
EDisCS paper. (See Sect.~\ref{sec:observations} and Table~\ref{tab:tab1} for
further details.)  A weak-shear analysis of gravitational lensing by our
clusters, based on these same data, is presented in \citet{Cloweetal05}.

{\it Phase III: Deep Near-Infrared Photometry:} For all but two of this same
set of $20$ cluster fields, we carried out deep near-infrared imaging using
SOFI at the NTT, obtaining at least $150$ minutes of exposure at K$_s$ for each
mid-$z$ cluster and at least $300$ minutes at J and $360$ minutes at K$_s$ for
each hi-$z$ cluster. The SOFI field is $5\farcm5\times 5\farcm5$ with a pixel
size of $0\farcs29$. The field of view of our SOFI images, taking account of
the dithering and overlapping of exposures, is $6.0^{\prime} \times
4.2^{\prime}$ for our mid-$z$ clusters and $5.4^{\prime} \times 4.2^{\prime}$
for our hi-$z$ clusters; see Arag\'on-Salamanca et al. (2005, in prep.)  for
details. Poor weather prevented observation of the last two of our mid-$z$
fields. These data allow us to characterise better the spectral energy
distributions, stellar masses and sizes of cluster members, and to obtain
reliable photometric redshifts for eliminating many foreground and background
galaxies.  They will be presented in detail in Arag\'on-Salamanca et al. (2005,
in prep.). Our photometric redshifts will be presented in Pell\'o et al. (2005,
in prep.).

{\it Phase IV: Confirmation Spectroscopy:} We used the multi-object slit-mask
capability of FORS2 to obtain spectra with $30$ minutes (mid-$z$) or $60$
minutes (hi-$z$) exposure for the candidate brightest cluster member and for up
to $30$ other candidate cluster members in each photometrically confirmed
cluster field. Redshift histograms were then constructed to confirm the
presence of a redshift spike in the range expected. One field was rejected at
this stage because its redshift histogram suggested that the detected object
was a superposition of a number of groups spread out in redshift, with no
dominant cluster.

{\it Phase V: Deep Spectroscopy:} For $18$ of the $19$ remaining
spectroscopically confirmed clusters we next obtained deep spectroscopy using
$3$ to $5$ separate slit masks on FORS2 and exposure times of typically
$60$--$120$ minutes (mid-$z$) or $240$ minutes (hi-$z$). This resulted in
samples of between $10$ and $70$ members in each cluster with an average of
about 30.  For the $19$th cluster we were able only to get one further $20$
minute integration with the existing mask. The spectra are of sufficient
quality to obtain both stellar population and kinematic information (velocity
dispersions for E/S0 galaxies, rotation velocities from [OII]$\lambda3727$ for
big disk galaxies).  Our spectroscopic observation procedures are presented in
detail together with data for our first $5$ clusters in \citet{Hetal04}.  The
data for the remaining $15$ clusters will be presented in Milvang-Jensen et
al. (2005, in preparation).

{\it Phase VI: Wide-field Imaging:} In order to obtain information about the
large-scale structure in which our clusters are embedded, we have used the
Wide-Field Imager at the ESO/MPG $2.2$m telescope to obtain $\sim120$ minute
exposures of each of our fields at R and $\sim60$ minute exposures in V and I.
The larger-scale environments of our clusters will be studied using these data
in Clowe et al. (2005b, in prep.)

{\it Phase VII: HST/ACS Imaging:}  For $10$ of our highest redshift clusters
we have obtained mosaic images with one orbit exposure in the outer regions
and $5$ orbit coverage at cluster centre using the Advanced Camera for Surveys
on board the Hubble Space Telescope. These data are particularly useful, in
combination with our deep VLT photometry, for studying the morphology of our
cluster galaxies.  The data will be presented and the HST morphologies
analysed in Desai et al. (2005, in prep.).

At the time of writing, data taking is finished for all phases with the final
observations obtained in spring 2004. Data reduction is completed for phases I,
II, III, IV, VI and VII and is almost completed for phase V.  Thus while we do
not, at this stage, have final results or interpretation for many aspects of
the survey, we do have enough information from all phases to give a clear view
of the size and quality of the final dataset. In this paper we present our
optical photometry in detail and give basic information about the cluster
sample, reserving detailed discussion of our other observations and the
presentation of our main results for future papers in this series.

\section{Optical photometry with FORS2}
\label{sec:photometry}

\subsection{The observations}
\label{sec:observations}

Our deep optical imaging programme was carried out using FORS2 in direct
imaging mode on the VLT. Our goal was to obtain $120$ minutes of high image
quality integration in V, R, and I for every cluster in the final hi-$z$ sample
and $45$ minutes of integration in B, V, and I for every cluster in the mid-$z$
sample. These are the standard FORS2 filters. B, V and I are close
approximations to the \citet{B90} photometric system while the R filter is a
special filter for FORS2.\footnote{The FORS2 filter curves are given at
http://www.eso.org/instruments/fors/inst/Filters/curves.html} The
imaging was performed in two phases - a contiguous $5$ night period in visitor
mode between March $19$ and March $24$ 2001, and $50$ hours in service mode
taken between April and July 2001, and in January and February 2002.  Two of
the nights during the March 2001 observations were photometric and all $20$
clusters had $10$ minutes of integration, taken in two $5$ minute exposures, in
all three passbands during these nights.  These short exposures were used to
define our photometric zero-points, and all of our other images were then
scaled to these zero-points by comparing the flux in bright but unsaturated
stars.  A list of the final exposure times,
FWHM seeing measurements and $5\sigma$ limiting magnitudes (for a $1\farcs$
radius aperture) is given in Table~\ref{tab:tab1}.

\begin{table*}
  \caption{FORS2 optical imaging: filters, exposure times, final seeing
    and $5\sigma$ limiting magnitudes}
  \label{tab:tab1}
  \begin{center}
    \begin{tabular}{l c c c c c c c c c}
      \hline
      mid-$z$ clusters && I &&& V &&& B\\
      \hline
      &t$_{\rm exp}$&FWHM&$m_{\rm lim}$&t$_{\rm exp}$&FWHM&$m_{\rm
        lim}$&t$_{\rm exp}$&FWHM&$m_{\rm lim}$\\
      &[min]&[$''$]&&[min]&[$''$]&&[min]&[$''$]\\
      \hline
      1018.8$-$1211 & 60 & 0.77 &24.9& 60 & 0.80 &26.1& 45 & 0.80 &26.5\\
      1059.2$-$1253 & 60 & 0.75 &24.7& 60 & 0.85 &26.2& 45 & 0.80 &26.4\\
      1119.3$-$1129 & 45 & 0.58 &24.8& 45 & 0.52 &26.2& 40 & 0.58 &26.5\\
      1202.7$-$1224 & 45 & 0.64 &24.9& 45 & 0.75 &26.0& 45 & 0.70 &26.4\\
      1232.5$-$1250 & 45 & 0.52 &24.8& 50 & 0.63 &26.2& 45 & 0.56 &26.4\\
      1238.5$-$1144 & 45 & 0.54 &24.8& 45 & 0.63 &26.1& 45 & 0.61 &26.6\\
      1301.7$-$1139 & 45 & 0.56 &24.9& 45 & 0.64 &26.2& 45 & 0.58 &26.4\\
      1353.0$-$1137 & 45 & 0.58 &25.0& 45 & 0.68 &26.1& 45 & 0.60 &26.2\\
      1411.1$-$1148 & 45 & 0.48 &25.0& 45 & 0.60 &26.2& 45 & 0.59 &26.4\\
      1420.3$-$1236 & 45 & 0.74 &24.9& 45 & 0.80 &26.1& 45 & 0.62 &26.4\\
      \hline
      hi-$z$ clusters && I &&& R &&& V\\
      \hline
      &t$_{\rm exp}$&FWHM&$m_{\rm lim}$&t$_{\rm exp}$&FWHM&m$_{\rm
        lim}$&t$_{\rm exp}$&FWHM&$m_{\rm lim}$\\
      &[min]&[$''$]&&[min]&[$''$]&&[min]&[$''$]\\
      \hline
      1037.9$-$1243 & 120 & 0.56 &25.1& 130 & 0.54 &26.1& 120 & 0.55 &26.6\\
      1040.7$-$1155 & 115 & 0.62 &25.1& 120 & 0.72 &26.0& 120 & 0.62 &26.5\\
      1054.4$-$1146 & 115 & 0.72 &25.1& 150 & 0.78 &26.1& 120 & 0.67 &26.6\\
      1054.7$-$1245 & 115 & 0.50 &25.4& 110 & 0.77 &26.1& 120 & 0.79 &26.5\\
      1103.7$-$1245 & 115 & 0.64 &25.3& 120 & 0.75 &26.1& 130 & 0.83 &26.4\\
      1122.9$-$1136 & 115 & 0.65 &25.2& 115 & 0.70 &25.9& 120 & 0.71 &26.3\\
      1138.2$-$1133 & 125 & 0.60 &25.2& 110 & 0.68 &26.0& 120 & 0.58 &26.5\\
      1216.8$-$1201 & 115 & 0.60 &25.3& 110 & 0.72 &26.0& 130 & 0.68 &26.6\\
      1227.9$-$1138 & 145 & 0.74 &25.1& 130 & 0.83 &26.0& 155 & 0.73 &26.4\\
      1354.2$-$1230 & 115 & 0.66 &25.2& 120 & 0.70 &26.0& 125 & 0.70 &26.4\\
      \hline
    \end{tabular}
  \end{center}
  Note: $m_{\rm lim}$ are $5\sigma$ limiting magnitudes in a $r=1\farcs0$
    aperture on our images convolved to $0\farcs 8$ seeing as estimated using
    the empty aperture simulations described in the text.
\end{table*}

\subsection{Production of calibrated images}
\label{sec:calibimages}

The images were reduced with the following steps. First a master bias frame
was scaled row-by-row to a linear 1-D fit to the over-scan strip of each image
and subtracted.  A $3\sigma$ clipped, medianed night-sky flat was then created
for each night in each band (or over contiguous nights if only a single field
was obtained on a given night in the service mode observations) and was used
to flat-field the images.  A bad pixel map was created from lists of pixels
either with exceptionally bright bias levels or with very low values in the
normalised flat-field.  Next, the sky was fit and subtracted using two
different methods, resulting in two sets of sky-subtracted data.  For both
methods, the IMCAT\footnote{ See
http://www.ifa.hawaii.edu/$\sim$kaiser/imcat.} {\it findpeaks} peak finder was
inverted to detect local minima in images smoothed with a $1\farcs$ Gaussian
to create a catalogue of local sky measurements.  The first method fit a
bi-cubic polynomial to these sky measurements, and subtracted the polynomial
from the unsmoothed image.  The second method used a $6\farcs5$ Gaussian
interpolation and smoothing method to create a sky-image which was subtracted
from the unsmoothed image.  This higher resolution sky-image corrected
small-scale variations in the sky level, but also subtracted the extended
wings of saturated stars and of the larger, foreground galaxies, as well as
any intergalactic light which might be present in the clusters.

An additional sky-subtraction step was deemed necessary as the I images had a
residual slope, after flat-fielding, along the east-west axis which correlated
with the total sky level.  Comparison of the photometry of objects in the same
field among images with differing sky levels (and therefore differing slopes)
did not reveal a similar slope in the photometry.  We therefore subtracted a
linear fit to this sky variation in order to remove it prior to combining
images. This minimises differences in sky level across the field.  The images
with high and low resolution sky subtraction were used to produce catalogues
for different purposes as noted below.

The IMCAT {\it findpeaks} peakfinder was used on both sets of sky-subtracted
data to detect bright but unsaturated stars.  The positions of these stars
were then compared to the USNO catalogue and to the positions in the other
exposures of the same field to derive both a correction for camera distortion
and the linear offset between the fields.  Each image was then distortion
corrected and aligned using a triangular method which preserves surface
brightness to map the images onto a common reference frame for each field.
This resulted in an alignment of the individual images to each other with a
stellar {\it rms} position error of $\sim 0.05$ pixels ($\sim 0\farcs 01$),
and an {\it rms} position offset between the images and the USNO catalogue
which varied from $0\farcs4$ to $0\farcs 6$ among the $20$ fields.  A final
image was then created using a sky-noise weighted mean with a clipping
algorithm which removed a pixel from the mean calculation if its value was
more than $3\sigma$ from the median {\it and} differed from the median by more
than $50$ per cent of the difference between the median and the sky level. The
second condition prevents clipping the centres of stars in images with
slightly different {\it psf}'s than the median.

The images created with the polynomial-fit sky-subtraction are used for our
photometric analysis and are the images on which the rest of this paper is
based.  The images created with the Gaussian-interpolated sky-subtraction are
used in our weak lensing analysis of the clusters, which is presented in Clowe
et al. (2005a).  A third set of images, created with the polynomial fit
sky-subtraction but without any distortion correction during the mapping, were
used to determine slit positions for our spectroscopy.  No colour-term
correction was applied to put these images on the standard BVRI system because
these corrections are very small for the FORS2 filter/CCD combination, smaller
than the errors on our photometric zeropoints. (From the dispersion in the
values we obtain for our standards, we estimate these zeropoint errors to be
0.018, 0.013, 0.013 and 0.016 magnitudes in B, V, R and I, respectively.)

\subsection{Catalogue construction}
\label{sec:catalogue}

In order to produce optical/IR photometric catalogues which are as uniform as
possible, we decided to match the seeing FWHM on all our optical and IR images
to $0\farcs 8$. (This was the typical seeing in our IR images; see
Arag\'on-Salamanca et al. (2005, in prep.).) This matching was done wherever
needed {\it and} possible. The few images with FWHM significantly larger than
$0\farcs 8$ were left untouched.  The following steps were used: (1) I-band
images were degraded to $0\farcs 8$ when necessary using a Gaussian convolution
kernel (IRAF/GAUSS), and (2) images in the other filters were similarly
degraded until their ``plumes'' of stars in a size versus magnitude plot
overlapped with the I-band plumes. The plumes were very well-defined, and this
made the matches quite straightforward.

Photometry on these seeing-matched images was carried out using SExtractor
version 2.2.2 \citep{BA96}. SExtractor was run in ``two-image'' mode using the
I-band images as detection reference images. Note that in cases where the
cluster I-band images needed to be degraded to $0\farcs 8$, the {\it original}
I-band images were used as the detection reference images. Thus our catalogues
are all based on detection in I-band at the original observed resolution, which
was often better than $0\farcs 8$, but our magnitudes (including those in
I-band) are measured whenever possible from images with FWHM of $0\farcs 8$.
Our photometric catalogues contain five types of I-band magnitude -- magnitudes
in circular apertures of radius $1$, $2$ and $3\farcs$, isophotal magnitudes
and SExtractor Kron (i.e. MAG--AUTO) magnitudes. All isophotal magnitudes for
each galaxy are measured within the {\it same} pixel set, defined to correspond
to the I-band isophote which is $1.5\sigma$ above the sky background. Similarly
our Kron magnitudes in all bands are measured within the aperture which
SExtractor defines for each galaxy based on its I-band image. We estimate an
approximate ``total'' I-magnitude for each galaxy by adding to the Kron
magnitude the correction appropriate for a point source measured within an
aperture equal to the galaxy's Kron aperture.  This correction is obtained from
an empirical ``curve of growth'' constructed from the images of bright,
unsaturated and isolated stars on each convolved I-band image.  It will
underestimate the correction needed obtain realistic {\it total} magnitudes,
particularly for large galaxies. At ${\rm I}_{\rm tot}\sim 25$ the median
correction from Kron to ``total'' magnitude ranges from 0.06 to 0.10 magnitudes
for our clusters. At these limits the median Kron magnitudes are themselves up
to 0.5 magnitudes brighter than the median isophotal magnitudes and up to 0.2
magnitudes brighter than the median magnitude within a $1\farcs$ aperture.

\begin{figure}
  \begin{center}
    \resizebox{\hsize}{!}{\includegraphics{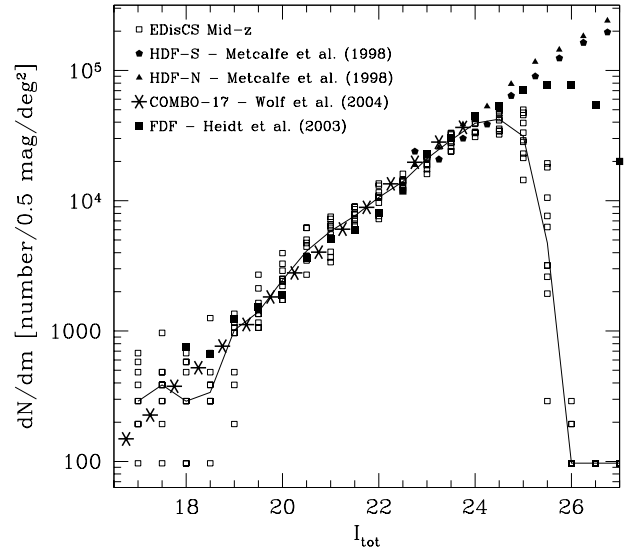}}
    \resizebox{\hsize}{!}{\includegraphics{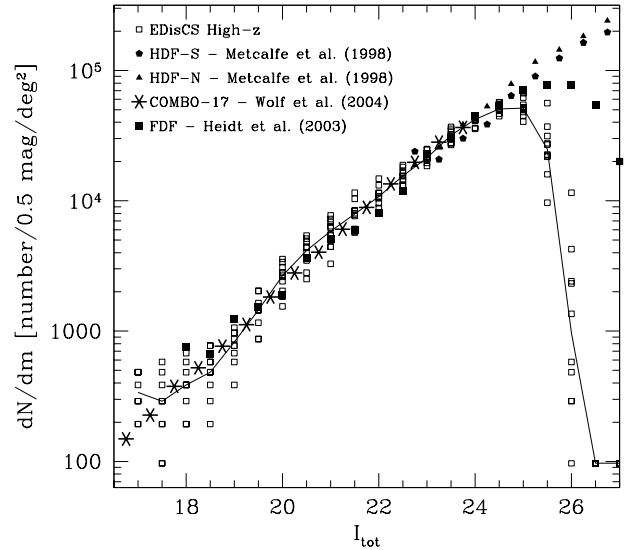}}
    \caption{Differential number counts in the I-band for all of our
      fields and separately for the mid-$z$ and hi-$z$ samples. Counts are
      given as the number per square degree per $0.5$ magnitude bin. For
      comparison we also show deeper counts for the two Hubble Deep Fields
      (Metcalfe et al. 2001) and for the FORS Deep Field (Heidt et al.
      2003) as well as shallower counts over a wider area
      from the COMBO-17 survey (private communication, see Wolf et al.
      (2003)). The thin solid line connects median count values for the
      ten EDisCS fields in each plot.}
    \label{fig:fig1}
  \end{center}
\end{figure}

Throughout this paper we correct our magnitudes and colours for Galactic
extinction according to the maps of \citet{SFD98} and a standard Milky Way
reddening curve.  We used the galactic extinction curve to derive the expected
extinction at the central wavelength of each of our filters. The reddening
value at each cluster's position is listed in Table~\ref{tab:tab2}. We made no
attempt to estimate or correct for differential reddening across our fields
Our catalogues, released with this paper, list data, uncorrected for Galactic
extinction, for all objects with ``total'' I-band magnitudes, defined as above
using the convolved images, brighter than ${\rm I} =24.0$ in the mid-$z$
fields and brighter than ${\rm I} =24.5$ in the deeper hi-$z$ fields (cf
Figure~\ref{fig:fig1}).

Our catalogues are available in electronic form at the CDS. The tables provide
64 columns containing: the galaxy ID; the object RA and Dec (J2000); five types
of magnitude (1,2,3 arcsec radius aperture magnitudes as well as isophotal,
SExtractor {\small MAG AUTO} magnitudes) and their corresponding errors for the
B, V, I (mid-$z$) and V, R, I (hi-$z$) bands; a point source aperture
correction; the ``total'' I magnitude ({\small MAG AUTO}+correction) and its
corresponding error; the SExtractor object flag in B, V, I (mid-$z$) and V, R,
I (hi-$z$) bands; the SExtractor Stellarity index in the same bands; the
SExtractor half-light radius in these bands; the SExtractor object ellipticity;
the SExtractor object position angle; the SExtractor object semi-major axis;
the SExtractor object semi-minor axis; the SExtractor object Kron radius; and
the SExtractor object isophotal area.

\subsection{Catalogue completeness}
\label{sec:completeness}

Understanding the completeness of our photometric catalogues is important for
many of our applications, but is far from simple.  Objects appear in the
catalogues if SExtractor detects them in our original I-band images, even
though the various magnitudes we quote are measured from the seeing-matched
images. Our completeness is high for galaxies whose I-band magnitude within a
$1\farcs$ radius aperture (after convolution to our fiducial seeing) is above
the $5\sigma$ limiting magnitude given in Table~\ref{tab:tab1}.  Such objects
are normally lost only if they become confused with other images. On the other
hand, objects with significantly brighter {\it total} magnitudes may be lost if
they have large sizes. In Fig.~\ref{fig:fig1} we show differential number
counts in the I-band for all of our fields. Star images have been removed from
these counts based on colour information and on the size-magnitude relation for
unresolved objects. The counts turn over sharply at the same point in all
fields in each set and at ``total'' I magnitudes comparable to the $5\sigma$
limits of Table~\ref{tab:tab1}.  This is a clear sign of incompleteness. The
scatter between the counts in different fields is larger than the counting
errors, demonstrating the well-known result that large-scale structure very
substantially affects counting statistics in fields of the size studied here
($46$ squ.arcmin.).  Comparison with deeper I-band counts from the two Hubble
Deep Fields \citep{Metc98} and from the FORS Deep Field \citep{Heidt2003} as
well as with somewhat shallower counts over a much larger area from the
COMBO-17 survey \citep{Wolfetal2003} shows very good agreement. Our
completeness clearly remains high to magnitudes significantly fainter than
${\rm I} =24$.

\subsection{Photometric errors}
\label{sec:errors}

Accurate photometric errors for galaxy colours are extremely important for
reliable photometric redshift determinations. Realistic errors for galaxy
magnitudes are also important for many aspects of the analysis we will carry
out in later papers. 

Our aperture and isophotal magnitudes are not intended as estimates of the
total light of our galaxies, but rather of the light in their images within the
chosen aperture after convolution to our fiducial seeing with FWHM $0\farcs 8$.
The uncertainty in such quantities is typically much less than in total
magnitudes for which the major source of scatter is a seeing-, model- and
object-dependent ``aperture correction''. This is why galaxy colours are much
more reliably estimated by differencing aperture or isophotal magnitudes than
by differencing total magnitudes. (Note that the isophote radius must be
defined in one band, the I-band in our case, and then used consistently for all
filters. Note also that the ``total'' magnitudes discussed above and plotted in
Fig.~\ref{fig:fig1} are not intended as accurate measures of total light, but
rather as robust lower limits which should approximate the true total light for
small and faint galaxies.) The errors in such colours should not depend on
galaxy structure or on seeing, but only on photon noise. For our convolved
images the photon noise in neighbouring pixels is correlated and is dominated
by sky counts for the great majority of objects. We therefore estimate this
noise as the {\it rms} scatter in the photon count over many apertures of each
relevant size placed at random in areas of each field where SExtractor detects
no object \citep[for details see][]{Letal03}. Errors in colours can then be
estimated as the sum in quadrature of the errors in the two relevant aperture
magnitudes.  This method ignores the possible contribution from systematic
effects which correlate between different bands observed with the same
instrument, as can arise, for example, from background sources which fall
beneath our detection threshold.

Although errors in our fixed aperture magnitudes (e.g. our standard 1, 2 and
$3\farcs$ apertures) are correctly estimated in this way, there is an
additional source of error in our isophotal, Kron and ``total'' magnitudes
(though not in colours derived from them) which comes from the uncertainty in
estimating the size of the isophote (or the Kron radius). Such errors can only
be estimated by careful Monte Carlo experiments in which artificial
``galaxies'' are inserted into real images and their properties measured just
as for the real galaxies. They depend on the exact magnitude definition and on
the details of the photometric software. We will report on such Monte Carlo
experiments for our EDisCS data in future papers where we require accurate
error estimates for specific applications.

\begin{figure}
  \begin{center}
    \resizebox{\hsize}{!}{\includegraphics{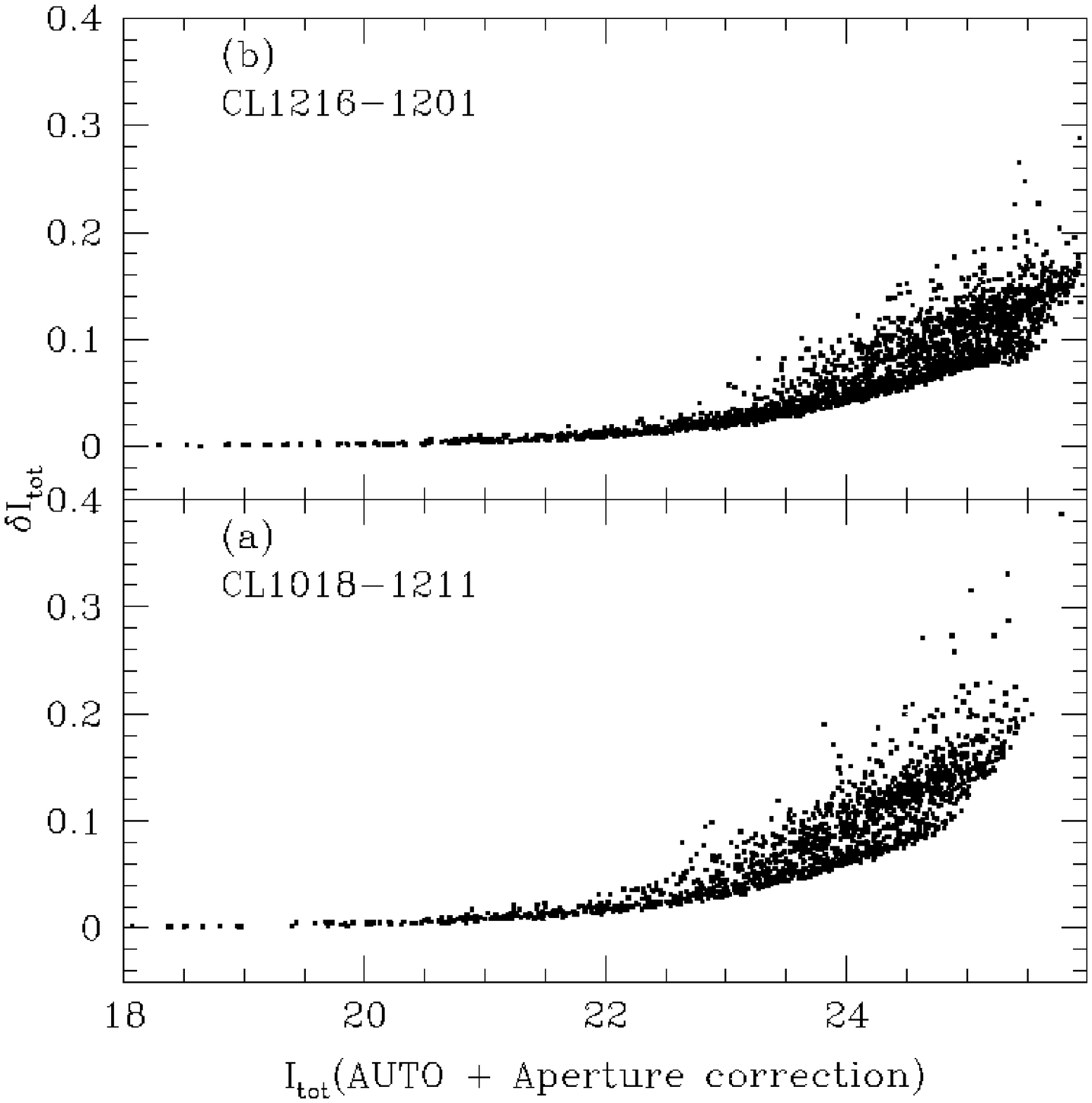}}
    \resizebox{\hsize}{!}{\includegraphics{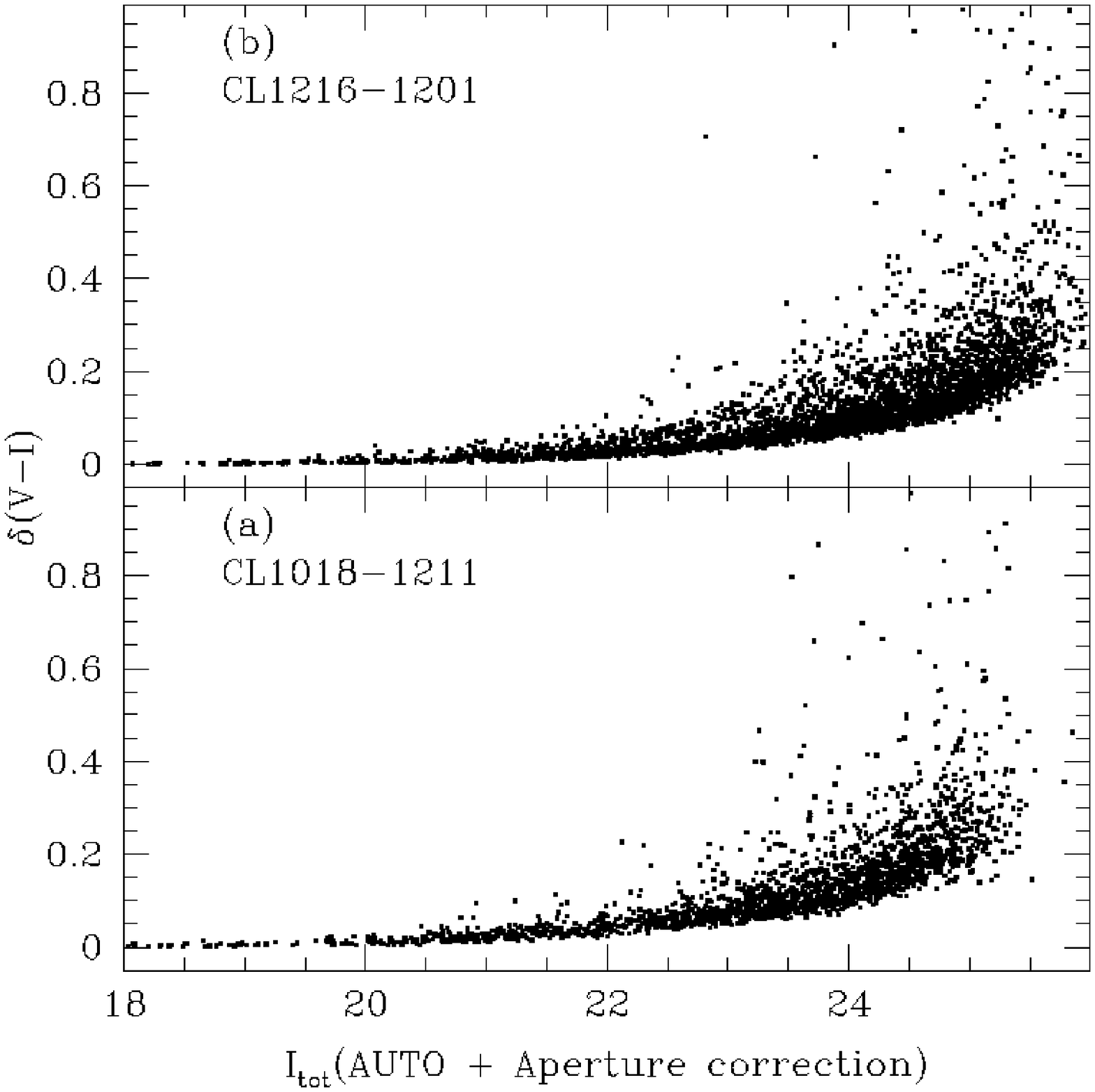}}
    \caption{Errors in our ``total'' I-band magnitudes and in our 
      ${\rm V}-{\rm I}$ colours are shown as a function of ``total'' I-band
      magnitude for typical fields from our mid-$z$ and hi-$z$ samples. These
      values were derived from Monte Carlo evaluation of the background count
      variance in apertures of the size used to measure the actual galaxy
      magnitudes (see text for details).}
    \label{fig:fig2}
  \end{center}
\end{figure}

We illustrate our error estimates in Fig.~\ref{fig:fig2}, which gives scatter
plots of the error in our I-band ``total'' magnitudes and in our ${\rm V}-{\rm
I}$ colours as a function of ``total'' magnitude for typical mid-$z$ and
hi-$z$ fields.  Note that in this and subsequent figures we shorten slightly
the names of the clusters to save space.  The colours for this plot are
computed using either the $1\farcs$ or the isophotal apertures for objects
depending on whether they are or are not flagged as being crowded by
SExtractor.  As just discussed, these are the errors due to sky noise in the
relevant apertures as estimated from the convolved images. They should thus be
accurate for the colour but somewhat too small for the ``total'' magnitude
because of the failure to include the uncertainty in the measured Kron radius
itself.  These errors are small all the way down to our estimated completeness
limit (see \S~3.4).  This can be understood by recalling that SExtractor
requires at least 4 adjacent pixels to be at least 1.5 sigma above the
estimated sky background in the filtered SExtractor detection image.
Particularly for extended objects this is a stringent requirement and causes
objects to drop out of the sample when the errors in their AUTO magnitudes are
small.

\section{The EDisCS Cluster Sample}
\label{sec:sample}

In this section we give basic information to characterise our sample of
clusters as a whole, together with a short description of each cluster.
Fig.~\ref{fig:fig3} presents false colour images of a square field $3\arcmin$
on a side centred on the galaxy which we identify as the brightest member of
each cluster (BCG; see the discussion of individual clusters below for detail
on how the BCG's were selected). These were constructed by combining the three
available optical images using a stretch which maximises the variation of image
colour with galaxy spectral energy distribution and so makes it easy to
recognise galaxies with similar colours, in particular red sequence galaxies
near the redshift of each cluster \citep[see][]{Letal04}. The colours and
luminosities of the BCG candidates themselves are plotted against redshift in
Fig.~\ref{fig:fig4} and are compared with a \citet{BC03} model for evolution
from a single, solar metallicity burst of star formation at redshift $3$. The
luminosity of the burst is normalised to give M$_{\rm V}=-22.4$ at $z=0$; this
is the absolute magnitude of NGC$4889$, the brightest galaxy in the Coma
Cluster.  The colours and the magnitudes of all our BCG candidates are
consistent with this passive evolution model.  

In Fig.~\ref{fig:fig5} we show I versus ${\rm V}-{\rm I}$ colour magnitude
diagrams within a $1.0$ Mpc radius circle centred on the BCG of each
cluster.\footnote{Throughout this paper we assume a flat cosmology with
$\Omega_{\rm m}=0.3$ and $H_0 = 70\,{\rm km}\,{\rm s}^{-1}\, {\rm Mpc}^{-1}$
whenever we quote quantities in physical rather than observed units.} These
were made using I-band ``total'' magnitudes and colours within $1\farcs$
radius apertures. We reject all galaxies for which our photometric redshift
programmes give a low probability of cluster membership. We derive photometric
redshifts from our optical + near-IR photometry using two independent codes
\citep{BMP00,Retal01,Retal03} both calibrated using our own spectroscopy; see
Pell\'o et al. (2005, in prep.) for details. Briefly we accept galaxies as
potential cluster members if the integrated photometric redshift probability
to be within $z\pm0.1$ of the cluster redshift is greater than a specific
threshold for {\it both} of the photometric redshift codes.  These
probability thresholds were calibrated using our spectroscopy and range from
0.1-0.35 depending on the filter set available for each particular field. No
further statistical correction for residual field contamination has been made
because the field of our infrared photometry does not extend much beyond the
region used to make the plots. We will come back to the issue of residual
contamination in later work.  Objects with sizes consistent with those of
unresolved point sources and spectral energy distributions consistent with
those of faint Galactic stars are rejected. The population of galaxies along
the red sequence of some of these clusters is analysed in \citet{DLGetal04}.

The galaxy we identify as the candidate BCG is shown in each colour-magnitude
diagram by a large cross.  The almost horizontal solid line gives the position
of the red sequence predicted by fitting the above Bruzual \& Charlot model to
the Coma cluster assuming that the slope reflects a dependence of metallicity
on stellar mass (i.e. fixing $z_{\rm burst} = 3$ for all masses). Parallel
dashed lines at $\pm 0.2$ magnitudes in colour delineate the band we use to
estimate a characteristic richness for each cluster. We define this as the
count of galaxies in this band brighter than the apparent magnitude which
translates to M$_{\rm V}=-18.2$ when evolved passively to $z=0$ with our
standard model. This apparent magnitude is shown as a vertical solid line in
each plot. It was chosen so that all objects counted are above the $5\sigma$
detection limit in all of our clusters with the exception of the most distant
object (cl1103.7$-$1245) which was identified late (see below). Finally, the
steep slanting dashed lines in each plot correspond to our $5$, $3$ and
$1\sigma$ detection limits in the V-band. Note that our richness measure is
not corrected for residual contamination. We estimate how large this may be as
follows. Our photo-$z$ rejection eliminates between 10 and 30 galaxies (median
18) from the bands of the colour-magnitude plots used for our richness count.
We have looked at the count distribution for colour-magnitude bands in random
patches of one field of the Canada-France Deep Field Survey
\citep{McCracken01} whose C-M position and area are matched to each of our
clusters in turn. The median count in these field patches ranges from 25 to 33
for the 20 clusters, but the count in different fields matched to the same
cluster ranges from below 10 to over 50, with a typical interquartile range of
a factor of 1.6.  Thus our richness count is probably biased high by about 10
galaxies, but the large expected fluctuation in background count makes any
case-by-case correction for residual contamination extremely uncertain.

In Fig.~\ref{fig:fig6} we give isopleths of the density of galaxies brighter
than I $=25$ in $4\farcm8\times 4\farcm8$ fields centred on our BCG candidates
after rejection of probable non-members and stars as above. Our cuts reject
$80$ to $90$ per cent of all objects to this magnitude limit, while rejecting
fewer than $10$ per cent of our spectroscopically confirmed cluster members.
This region is the largest square field for which we have both IR and optical
data for most of our clusters. These contours were made by adaptive kernel
smoothing in a scheme very similar to that outlined by \citet{P93,P96}.  The
position of our BCG candidate in each field is marked by a cross. An artifact
should be noted in these plots; the densities always drop close to the field
boundary because we have not attempted to correct for boundary effects. Note
also that for two clusters we do not have IR data and in these fields our
photometric redshift techniques are less efficient at rejecting non-members (cf
Table~\ref{tab:tab2}).

Numerical data for each cluster are given in Table~\ref{tab:tab2} which lists
the cluster name, the position, its original ID in the LCDCS, the reddening
value adopted for each field, I-band ``total'' apparent magnitude, ${\rm
  V}-{\rm I}$ colour (both corrected for Galactic extinction) and redshift of
the BCG, as well as the measure of cluster richness we discussed above. We
reiterate that this is effectively a count of ``red sequence'' galaxies above a
fixed stellar mass limit, so that its relation to the total richness of a
cluster depends on the relative proportions of red and blue galaxies.  As may
be seen from Fig.~\ref{fig:fig5}, our sample shows large variations in this
ratio. Some clusters appear to show a strong red sequence with rather few blue
galaxies (e.g.  cl1232.5$-$1250, cl1138.2$-$1133) while others show many blue
galaxies but relatively few passively evolving systems (e.g. cl1238.5$-$1144,
cl1040.7$-$1155, cl1227.9$-$1138).  This is an interesting result which may in
part reflect differences in foreground/background contamination. We will come
back to it again in later papers.

\begin{table*}
\caption{Cluster Data}
\label{tab:tab2}
\begin{center}
\begin{tabular}{l c c c c c c c c}
  \hline
  Cluster & BCG Coordinates$^{\rm a}$ & LCDCS ID  & E(B-V) & I$^{\rm b}$ &  V -I $^{\rm c}$ & $z$   & Richness$^{\rm d}$  & Filters\\  
  \hline
1018.8$-$1211 &10:18:46.8 -12:11:53 & 0057 &0.077& 17.96 & 2.16 & 0.47        & 46   & BVIK$_{s}$\\
1059.2$-$1253 &10:59:07.1 -12:53:15 & 0188 &0.033& 17.84 & 2.14 & 0.46$^{\rm e}$& 51 & BVIK$_{s}$\\
1119.3$-$1129 &11:19:16.7 -11:30:29 & 0252 &0.066& 19.24 & 2.18 & 0.55        & 28   & BVI\\
1202.7$-$1224 &12:02:43.4 -12:24:30 & 0430 &0.058& 18.40 & 1.99 & 0.42        & 39   & BVIK$_{s}$\\
1232.5$-$1250 &12:32:30.5 -12:50:36 & 0541 &0.060& 18.20 & 2.36 & 0.54        & 105  & BVIJK$_{s}$\\
1238.5$-$1144 &12:38:33.0 -11:44:30 & 0567 &0.044& 18.37 & 2.19 & 0.46        & 32   & BVI\\
1301.7$-$1139 &13:01:40.1 -11:39:23 & 0634 &0.049& 18.27 & 2.21 & 0.48        & 48   & BVIK$_{s}$\\
1353.0$-$1137 &13:53:01.7 -11:37:28 & 0849 &0.063& 18.64 & 2.44 & 0.59        & 28   & BVIK$_{s}$\\
1411.1$-$1148 &14:11:04.6 -11:48:29 & 0925 &0.065& 18.25 & 2.24 & 0.52        & 45   & BVIK$_{s}$\\
1420.3$-$1236 &14:20:20.0 -12:36:30 & 0952 &0.082& 18.53 & 2.24 & 0.50        & 37   & BVIK$_{s}$\\
1037.9$-$1243 &10:37:51.2 -12:43:27 & 0110 &0.043& 18.54 & 2.29 & 0.58$^{\rm e}$& 22 & VRIJK$_{s}$\\
1040.7$-$1155 &10:40:40.4 -11:56:04 & 0130 &0.051& 19.87 & 2.49 & 0.70        & 23   & VRIJK$_{s}$\\
1054.4$-$1146 &10:54:24.5 -11:46:20 & 0172 &0.037& 19.49 & 2.56 & 0.70        & 53   & VRIJK$_{s}$\\
1054.7$-$1245 &10:54:43.6 -12:45:52 & 0173 &0.038& 19.61 & 2.68 & 0.75        & 33   & VRIJK$_{s}$\\
1103.7$-$1245 &11:03:43.4 -12:45:34 & 0198 &0.048& 20.68 & 2.73 & 0.96        &15  & VRIJK$_{s}$\\
1103.7$-$1245a &11:03:34.9 -12:46:46 &     &0.048& 19.54 & 2.46 & 0.63        && VRIJK$_{s}$\\
1103.7$-$1245b &11:03:36.5 -12:44:22 &     &0.048& 19.46 & 2.57 & 0.70        &  & VRIJK$_{s}$\\

1122.9$-$1136 &11:22:51.6 -11:36:32 & 0275 &0.040& 19.50 & 2.23 & 0.64        & 15   & VRIJK$_{s}$\\
1138.2$-$1133 &11:38:10.3 -11:33:38 & 0340 &0.027& 18.63 & 2.09 & 0.48        & 49   & VRIJK$_{s}$\\
1216.8$-$1201 &12:16:45.1 -12:01:18 & 0504 &0.045& 19.46 & 2.63 & 0.80        & 74   & VRIJK$_{s}$\\
1227.9$-$1138 &12:27:58.9 -11:35:13 & 0531 &0.047& 19.33 & 2.41 & 0.63        & 28   & VRIJK$_{s}$\\
1354.2$-$1230 &13:54:09.7 -12:31:01 & 0853 &0.079& 19.47 & 2.53 & 0.76        & 22   & VRIJK$_{s}$\\
    \hline
\end{tabular}\\
$^{\rm a}$ Coordinates (J2000) of our BCG candidate\\
$^{\rm b}$ BCG ``total'' magnitude from SExtractor AUTO-MAG + a minimal 
point source correction (always $<0.04$mag for these galaxies)\\
$^{\rm c}$ BCG colour measured within a $r=1\farcs$ circular aperture\\
$^{\rm d}$ Number of red-sequence galaxies within 1 Mpc of the BCG (after
  photo-z rejection of non-members)\\and brighter
  than the apparent I magnitude which evolves passively to M$_{\rm V}=-18.2$ at
  $z=0$\\ 
$^{\rm e}$ Redshift of the cluster -- no spectroscopy for BCG\\
\end{center}

\end{table*}

Our clusters were originally taken from the LCDCS lists.  In many fields the
identification of the cluster and the associated BCG is unambiguous, based on
the available spectroscopy, the brightness, colour, and spatial distributions
of galaxies in the images, their photometric redshifts and our weak lensing
maps.  Nevertheless, there were some fields where it was not possible to
conclusively identify the BCG.  We will not discuss the spectroscopic
redshifts, lensing maps, or photometric redshifts in detail here, since these
will be presented elsewhere (Halliday et al. 2004; Clowe et al. 2005a;
Milvang-Jensen et al. 2005, Clowe et al. 2005b, Pell\'o et al. 2005, all in
prep.).  We do, however, give a description of each cluster based on all
currently available information. This includes the basis for our current
identification of the BCG and redshift of the cluster, together with our
current assessment of the structure and richness of the cluster based on the
images and colour-magnitude diagrams of Figs.~\ref{fig:fig3} and
\ref{fig:fig5}.  The spectroscopic data for $5$ of our clusters
(cl1232.5$-$1250, cl1040.7$-$1155, cl1054.4$-$1146, cl1054.7$-$1245, and
cl1216.8$-$1201) were presented in \citep{Hetal04}.  The data for the remaining
$15$ clusters will be presented in Milvang-Jensen et al. (2005, in
preparation).

Finally we note that one field (cl1103.7$-$1245) has three clusters listed in
Table~\ref{tab:tab2}. As we discuss below, when we reduced our deep
spectroscopy for this field we discovered a distant cluster at the LCDCS
position but at a substantially higher redshift ($z=0.96$) than that inferred
from our confirmation spectroscopy and so built into our spectroscopic
targetting strategy. At the lower redshift two substantial groupings showed up
in the field but centred away from the LCDCS position.  Our BCG candidates for
all three structures are plotted in Fig.~\ref{fig:fig4}, and colour-magnitude
diagrams and isopleths are shown for the lower redshift range in the last
panels of Fig.~\ref{fig:fig5} and Fig.\ref{fig:fig6}, respectively, after
photo-$z$ rejection of foreground and background galaxies.

{\bf Mid-$z$ clusters}
 
\textit{cl1018.8$-$1211}: Redshifts are available for $69$ galaxies in this
field. In the corresponding spectroscopic redshift histogram there is a clear
peak of more than $30$ spectroscopically confirmed members at $z=0.47$.  There
is also a peak in the photometric redshift probability distribution at
$z\approx0.47$.  The isopleths of Fig.~\ref{fig:fig6} show a clear
concentration of galaxies around the cluster centre with a second
concentration about $3\arcmin$ to the southwest.  There is a $3.0\sigma$
detection in the weak lensing map centred on the concentration of red galaxies
in the image. The colour-magnitude diagram suggests a well defined red
sequence with a significant blue population.  The brightest galaxy at the
LCDCS position is at the centre of the galaxy distribution and has a redshift
in the spectroscopic peak.  We identify this galaxy, EDCSNJ1018467$-$1211527,
as the BCG.

\textit{cl1059.2$-$1253}: Redshifts are available for $86$ galaxies in this
field.  In the spectroscopic redshift histogram there is a clear peak of $40$
objects at $z=0.46$.  There is also a peak in the photometric redshift
probability distribution at $z\approx0.45$.  The isopleths show a very strong
peak at the LCDCS position. There is a $5.2\sigma$ detection in the weak
lensing map centred on this concentration of red galaxies.  The
colour-magnitude diagram demonstrates that this is a rich cluster with a
well-defined red sequence but also with many bluer galaxies. The brightest
galaxy at the LCDCS position is near the centre of the galaxy distribution, it
has a photometric redshift equal to $0.42$, but does not have a measured
spectroscopic redshift.  We identify this galaxy, EDCSNJ1059071$-$1253153, as
the BCG.

\textit{cl1119.3$-$1129}: Redshifts are available for $57$ galaxies in this
field. The redshift histogram has $22$ objects at $z=0.55$, and there is a
peak in the photometric redshift probability distribution at $z\approx0.55$.
The isopleths show no clear concentration of galaxies in the field and there
is no weak lensing detection. The colour-magnitude diagram shows a relatively
strong red sequence at the expected colour but bluer galaxies appear absent
(perhaps because the absence of K-band data affects our photometric redshift
probabilities in this field). The identification of the BCG is uncertain;
there are multiple bright galaxies in the vicinity of the LCDCS detection
which could be the BCG.  For now, we treat the object EDCSNJ1119168$-$1130290
as the BCG since we have a spectrum.  It is fainter than expected for a BCG,
however (see Table~\ref{tab:tab2} and Fig.~\ref{fig:fig4}) and there are
candidates with similar colours and brighter magnitudes in the field, but with
no spectroscopy.

\textit{cl1202.7$-$1224}: Redshifts are available for $62$ galaxies in this
field. The redshift histogram has a dominant
peak with more than $20$ objects at $z=0.42$.  There is also a peak in the
photometric redshift probability distribution at $z\approx0.42$.  The isopleths
in this field show a clear concentration near the expected position. There is a
$1.5\sigma$ detection in the weak lensing map centred on the concentration of
red galaxies.  The colour-magnitude diagram shows a clear red sequence at the
expected colour with a few bluer galaxies.  The brightest galaxy at LCDCS
position is quite close to the centre of the galaxy distribution and has a
redshift in the spectroscopic peak.  We identify this galaxy,
EDCSNJ1202433$-$1224301, as the BCG.

\textit{cl1232.5$-$1250}: Redshifts are available for $100$ galaxies in this
field.  This is one of our richest clusters with a peak of $54$ objects in the
spectroscopic redshift histogram at $z=0.54$.  There is also a peak in the
photometric redshift probability distribution at $z\approx0.54$.  The
isopleths show a strong and highly elongated concentration and there is a
$7.6\sigma$ detection in the weak lensing map centred on this concentration.
The colour-magnitude diagram shows a very strong red sequence but also a
substantial population of bluer galaxies.  The brightest galaxy near the LCDCS
position is close to the centre of the galaxy distribution.  We identify this
galaxy, EDCSNJ1232303$-$1250364, as the BCG.

\textit{cl1238.5$-$1144}: This cluster has the poorest spectroscopic coverage
in our sample, with only two short exposure masks obtained.  Redshifts are
available for only $10$ galaxies in the field.  Because of the obvious
weakness of this cluster it was assigned low priority for long-exposure
spectroscopy, and it is the only one of our $19$ retained clusters for which
we eventually obtained {\it no} longer exposure masks.  In the first mask
there are at least $4$ galaxies at $z=0.46$.  There is a broad peak in the
photometric probability distribution at $z\sim 0.45$.  The broadness of the
peak is probably due to the lack of NIR data in this field. There is a
concentration in the isopleths at the expected position but it is very weak.
Furthermore there is no weak lensing detection for this field.  The
colour-magnitude diagram shows a relatively strong red sequence at the
expected colour.  The brightest galaxy at the LCDCS position is close to the
peak of the isopleths and has a redshift in the spectroscopic peak.  We
identify this galaxy, EDCSNJ1238330$-$1144307, as the BCG.

\textit{cl1301.7$-$1139}: Redshifts are available for $85$ galaxies in this
field. The redshift histogram of this field has $37$ objects at $z=0.48$ and
$16$ objects at $z=0.39$. The coordinate of the LCDCS source matches that of
the brightest galaxy in the $0.48$ peak.  The photometric redshift probability
distribution has a peak at $z\approx0.48$, but nothing obvious at $0.40$.  The
isopleths in the field show two clear concentrations separated by about
$2\arcmin$.  The southeastern peak is the $z=0.48$ cluster targeted from the
LCDCS, the northwestern peak is likely the $0.4$ cluster.  There is a
spatially extended weak lensing detection at $3.4\sigma$ that overlaps with
the galaxy concentrations in the two peaks. The red sequence in the
colour-magnitude diagram is strong and appears concentrated at the colour
expected for the higher redshift.  There is also a substantial population of
blue galaxies.  The $0.48$ structure was the one selected by the LCDCS and we
choose the brightest galaxy in this concentration, EDCSNJ1301402$-$1139229, as
the BCG.

\textit{cl1353.0$-$1137}: Redshifts are available for $65$ objects in this
field. The redshift histogram has at least $27$ objects at $z=0.58$.  There is
also a peak in the photometric redshift probability histogram at
$z\approx0.58$. The isopleths are peaked at the expected position, but less
strongly than in many of our other clusters. A peak in the lensing map also
coincides with the obvious galaxy concentration on the image at the location
of the LCDCS source.  The colour-magnitude diagram shows a well-defined but
relatively poor red-sequence at the expected colour. There is a substantial
population of bluer galaxies.  The brightest galaxy at the position of the
LCDCS source is located at the centre of the galaxy distribution and has
$z=0.59$.  We identify this galaxy, EDCSNJ1353017$-$1137285, as the BCG.

\textit{cl1411.1$-$1148}: Redshifts are available for $73$ galaxies in this
field.  There are $23$ objects at $z=0.52$.  There is also a peak in the
photometric redshift probability histogram at $z\approx0.52$. The isopleths
show a strong and relatively symmetric concentration at the expected
position. There is a peak in the lensing map coincident with this
concentration which is also at the LCDCS location.  The colour magnitude
diagram shows a red sequence which appears slightly redder than
expected. There are fewer blue galaxies here than in many of our clusters. The
brightest galaxy at the LCDCS position has a redshift in the spectroscopic
peak.  We identify this galaxy, EDCSNJ1411047$-$1148287, as the BCG.

\textit{cl1420.3$-$1236}: Redshifts are available for $70$ galaxies in this
field. In the redshift histogram there are $26$ objects at $z=0.49$.  There is
also a peak in the photometric redshift probability histogram at
$z\approx0.50$.  The isopleths have a clear and symmetric peak at the expected
position, but although there is a $2.7\sigma$ peak in the lensing map, it is
not perfectly coincident with the galaxy concentration (which does coincide
with the LCDCS source). The colour-magnitude diagram for this field shows a
weak red sequence at the expected colour in addition to the population of
bluer galaxies. The brightest galaxy at the LCDCS position is near the peak of
the isopleths and has a redshift in the spectroscopic peak.  We identify this
galaxy, EDCSNJ1420201$-$1236297, as the BCG.

{\bf Hi-$z$ clusters}

\textit{cl1037.9$-$1243}: Redshifts are available for $117$ galaxies in this
field.  The spectroscopic redshift histogram for cl1037.9$-$1243 is complex
with two dominant peaks at $z=0.42$ ($43$ members) and $0.58$ ($19$ members).
The relative strengths of the peaks may be biased by our targetting brighter
galaxies for our initial spectroscopy.  There is an obvious cigar-shaped
cluster in this field at the position of the LCDCS source with a concentration
of red galaxies of similar colour.  While the objects in the $0.43$ spike are
spread over the whole field, the objects in the $0.58$ spike are concentrated
around the LCDCS source position.  There is a peak in the photometric redshift
probability distribution at $z\approx 0.58$, but our photo-$z$'s do not
perform well at $z\lesssim0.5$ for the hi-$z$ filter set (cf.
Table~\ref{tab:tab2}).  Objects with photo-$z$'s around $0.58$ are
concentrated at the position of the cigar while the objects with $z_{\rm
phot}$ consistent with $0.42$ are concentrated towards the west side of the
image. The isopleths for the field are quite complex with no obvious dominant
peak.  All the principal structures seen in the isopleths are also detected in
the weak lensing map.  In the colour-magnitude diagram there is no clear red
sequence at $z=0.58$ but a reasonably well populated one at $z\sim 0.42$
(compare with cl1202.7$-$1224).  Although there is obviously a rich cluster at
$0.42$, the object selected by the LCDCS nevertheless appears to be the
cluster at $z=0.58$.  Our original choice for the BCG was resolved into two
objects by our ACS image of this cluster.  The brighter of these may still
turn out to be the BCG, but there are also other bright objects in the cluster
core.  We provisionally select one of these, EDCSNJ1037514$-$1243266, as the
BCG of the $z=0.58$ structure.  For this galaxy, we do not have spectroscopic
information.

\textit{cl1040.7$-$1155}: Redshifts are available for $130$ galaxies in this
field.  In the spectroscopic redshift histogram there is a peak of $30$
objects at $z=0.70$.  There is also a peak in the photometric redshift
probability distribution at $z\approx 0.70$. The isopleths show a single
dominant concentration at the LCDCS position but no significant mass
concentration is detected in the weak lensing map.  The colour magnitude
relation shows a remarkably weak red sequence but a substantial population of
bluer galaxies.  The brightest galaxy near the LCDCS position is centred on
the peak of the isopleths and has a redshift in the spectroscopic peak.  We
identify this galaxy, EDCSNJ1040403$-$1156042, as the BCG.

\textit{cl1054.4$-$1146}: Redshifts are available for $109$ galaxies in this
field.  In the spectroscopic redshift histogram there are $49$ cluster members
at $z=0.70$.  There is a peak in the photometric redshift probability
distribution at $z\approx 0.70$.  The dominant peak in the isopleths is at the
LCDCS position which also coincides with a concentration in the weak lensing
mass map.  The colour-magnitude diagram shows a well developed red sequence at
the expected colour but also a population of bluer galaxies including some
bright objects.  The brightest galaxy near the LCDCS position is at the peak
of the isopleths and has a redshift in the spectroscopic peak.  We identify
this galaxy, EDCSNJ1054244$-$1146194, as the BCG, although there are other
possible candidates of similar colour and slightly brighter magnitude.

\textit{cl1054.7$-$1245}: Redshifts are available for $104$ galaxies in this
field. In the spectroscopic redshift histogram there are $36$ objects at
$z=0.75$, as well as a group of $10$ galaxies at $z=0.73$ which is about
$7\sigma$ away from the main cluster where $\sigma$ is the cluster velocity
dispersion \citep{Hetal04}.  There is a peak in the photometric redshift
probability distribution at $z\approx 0.75$.  The galaxy isopleths show a
strong peak at the LCDCS position and there is also a strong concentration in
the weak lensing mass at this position. This cluster has a well developed red
sequence which appears somewhat redder than the expected colour, together with
a substantial population of bluer galaxies, again including a number of bright
galaxies. As can be seen in Fig.~\ref{fig:fig3}, this cluster has one or two
strongly lensed arcs directly north of our BCG candidate. This is
EDCSNJ1054435$-$1245519, the brightest galaxy at the LCDCS position. It is
centred on the isopleth peak and has a redshift in the spectroscopic peak.

\textit{cl1103.7$-$1245}: Redshifts are available for $100$ galaxies in this
field. Our initial spectroscopy showed a number of galaxies in the redshift
range $0.6$ to $0.7$, so we used our photometric redshifts to reject objects
far from this redshift when selecting targets for deep spectroscopy.  The final
spectroscopic redshift histogram indeed shows peaks at $z=0.70$ ($11$ members)
and $0.63$ ($14$ members).  There is also, however, a peak containing nine
objects at $z=0.96$ despite our targetting bias against such high redshifts.
There are peaks in the photometric redshift probability distribution at $z\sim
0.65$, at $z\sim 0.70$ and also at $z\sim 0.95$.  The brightest galaxies at
$z=0.63$ and $z= 0.70$ are located quite far away from the LCDCS position, but
the brightest galaxy in the $0.96$ peak coincides with the LCDCS position and
with a bright field galaxy at $z=0.66$. We have made two sets of
colour-magnitude and isopleth plots, one rejecting galaxies with photo-$z$'s
incompatible with $z=0.96$, and the other rejecting galaxies incompatible with
$z=0.66$ (in order to include the structures at redshifts $0.63$ and $0.70$).
We show the plots for the $z=0.96$ case labelled \textit{cl1103$-$1245} in
figures 5 and 6, where the C-M diagram is for a circle radius 1.0 Mpc
surrounding the BCG, as for the other clusters.  The plots for the $z=0.66$
case are also shown, and are labelled \textit{cl1103$-$1245a/b}. The C-M
diagram here is for the full region for which we have five band photometry. The
isopleths at $z=0.96$ show a single peak close to the LCDCS position. At
$z=0.66$ the structure is more complex with several peaks. The weak lensing
maps indicate a $3\sigma$ detection near the LCDCS source, but also a detection
near the brightest galaxy at $z=0.63$, while we do not have any detection close
to the brightest galaxy at $z=0.7$.  There is also a grouping of red galaxies
around the brightest $z=0.63$ galaxy -- no such overdensity of red objects is
apparent around the brightest galaxy at $z=0.70$.  The colour-magnitude
diagrams show reasonably well-defined red sequences at the expected colours. We
consider the $z=0.96$ cluster at the LCDCS position as the selected cluster and
choose its brightest galaxy, EDCSNJ1103434$-$1245341 at $z=0.9576$, as BCG.
Both the $z=0.63$ and the $z=0.70$ concentrations also contain bright galaxies
with the visual and photometric properties of BCG's.  These are
EDCSNJ1103349$-$1246462 at $z=0.6257$ and EDCSNJ1103365$-$1244223 at
$z=0.7031$; we give their properties in Fig.~4 and Table~\ref{tab:tab2} where
they are denoted \textit{cl1103.7$-$1245a} and \textit{cl1103.7$-$1245b},
respectively.

\textit{cl1122.9$-$1136}: From the 2002 (short exposure) mask there was no
evidence of a concentration of galaxies at any specific redshift in this
field.  In addition, our photo-$z$'s show no clear peak and our weak lensing
map does not have a significant maximum. The isopleths do show a convincing
peak at the LCDCS position which coincides with a galaxy with $z=0.64$ with
approximately the colour and magnitude expected for a BCG. This is presumably
what the LCDCS detected.  On the other hand, our final colour-magnitude plot
shows no convincing red sequence at the expected colour.  After some
discussion we decided to drop this field from our spectroscopic programme,
thus no mask with long exposure was observed. Our initial candidate for BCG
was the galaxy EDCSNJ1122517$-$1136325.

\textit{cl1138.2$-$1133}: Redshifts are available for $106$ galaxies in this
field. Although this field was originally selected as one of our high redshift
clusters, the dominant peak in spectroscopic redshift histogram ($48$ galaxies)
is at $z=0.48$ and coincides spatially with the LCDCS source and the
concentration of red galaxies in the image.  (It turns out that the LCDCS
overestimated the redshift of this cluster because the cluster centre and the
true BCG lay behind the mask.) The isopleths show a single well defined peak
within an elongated filament-like structure. Although there are also $5$
galaxies at $z=0.62$, these are spread over the field.  Our photometric
redshifts are not robust at $z\lesssim0.5$ for the hi-$z$ filter set (cf.
Table~\ref{tab:tab2}) and we do not find a peak corresponding to the
spectroscopic peak.  There is a weak lensing concentration at the location 
of the LCDCS source.  The colour-magnitude diagram shows a tight red
sequence at bright magnitudes and at the colour expected for $z=0.48$. There
are surprisingly few blue galaxies or red sequence galaxies at fainter
magnitudes, but this might be related to the problem of estimating reliable
photometric redshifts with this $z$ and filter set.  The brightest galaxy at
the LCDCS position is centred on the isopleth peak and has a redshift in the
spectroscopic peak.  We identify this galaxy, EDCSNJ1138102$-$1133379, as the
BCG.

\textit{cl1216.8$-$1201}: This is one of our richest and highest redshift
clusters.  Redshifts are available for $120$ galaxies in the field. In the
spectroscopic redshift histogram there are $66$ objects at $z=0.79$.  There is
a peak in the photometric redshift distribution at the redshift of the cluster
and a $6.3\sigma$ weak lensing detection which overlaps with the galaxy
concentration.  There is also a strongly lensed arc in this cluster. The
isopleths show a strong peak at the expected position but also several
additional peaks which may be cluster substructure. The colour-magnitude
diagram shows a very well developed red sequence and also a well defined and
well populated sequence of (presumably) star-forming galaxies about $1.5$ mag
bluer.  The brightest galaxy at the LCDCS position is centred on the isopleth
peak and has a redshift in the spectroscopic peak.  We identify this galaxy,
EDCSNJ1216453$-$1201176, as the BCG.

\textit{cl1227.9$-$1138}: Redshifts are available for $113$ galaxies in this
field which has a complicated structure.  There is a dominant peak of $22$
objects at $z=0.63$.  There are, however, at least $2$ other spikes at $z=0.49$
and $0.83$, and a broad concentration near $0.58$.  The galaxies at $z=0.63$
coincide with the LCDCS detection, but so do those in the $z=0.49$ peak.  There
is also a peak in the photometric redshift probability distribution at
$z\approx 0.63$, but none at $z=0.49$ (perhaps because of the poor performance
of our photo-$z$ estimates for this filter set at $z\lesssim0.5$). The
strongest peak in the isopleths is indeed close to the LCDCS position but there
are $5$ or $6$ additional peaks in the field.  There is no clear weak lensing
detection associated with any of these.  The colour-magnitude diagram does show
a possible red sequence at $z=0.63$ but the brightest red galaxies appear to be
associated with a red sequence at $z=0.49$. There are many blue galaxies in the
colour-magnitude range expected for objects at either of these redshifts.  We
believe that the $z=0.63$ spike is the cluster picked by the LCDCS. The galaxy
EDCSNJ1227539$-$1138173 is the brightest object with a spectroscopic redshift
at $z=0.63$ near the centre of the red galaxy concentration and the LCDCS
position. Its magnitude, however, is significantly fainter than expected for a
BCG and the galaxy EDCSNJ1227589$-$1135135 about 2 arcmin to the north is about
$1.2$ magnitudes  brighter and has redshift $z=0.6375$. We provisionally choose
this latter object as the BCG, although in this case we still centre the false
colour image, the C-M diagram, the isopleths, and calculate the richness for a
1.0 Mpc circle surrounding EDCSNJ1227539$-$1138173 and the LCDCS position.

\textit{cl1354.2$-$1230}: Redshifts are available for $116$ galaxies in this
field.  In the spectroscopic redshift histograms there are $20$ objects at
$z=0.76$, and smaller peaks at $z=0.52$ and $0.59$.  The composite nature of
the spectroscopic redshift histogram may be a result of the wider than normal
photo-$z$ interval used to select galaxies for spectroscopy.  This wider
interval was used because the initial 2002 spectroscopy showed the secondary
spike at $z=0.6$ and we were unsure which was the dominant cluster.  There are
also peaks in the photometric redshift probability distribution at
$z\approx0.6$ and $0.7$. The isopleths in this field show a strong peak very
close to the LCDCS position but there are clearly a number of subsidiary peaks
in the field. There is a $3.9\sigma$ lensing detection centred on the dominant
peak which is a concentration of red galaxies. These show up in our
colour-magnitude diagram as a relatively poor red sequence at the expected
colour. The fainter red sequence is remarkably empty, but there are plenty of
blue cluster members at the expected colours and magnitudes. The brightest
galaxy near the main isopleth peak has a spectroscopic redshift of $z=0.76$,
and we identify this galaxy, EDCSNJ1354098$-$1231015, as the BCG of a cluster
at this redshift.

\section{Conclusion}
\label{sec:conclusion}

The sample of clusters we present here has been uniformly selected and
uniformly observed. Moreover, the quality of the data delivered by the ESO
telescopes is uniformly high, thanks to the superb instrumentation, the good
conditions at Paranal, and the possibility of obtaining data in service mode.
The estimated redshifts provided by the LCDCS scatter by $0.06$--$0.08$ {\it
rms} about the true values and are systematically high by about $0.1$ at high
redshift. As a result we have ended up with a set of clusters distributed
relatively smoothly in redshift between $0.42$ and $0.96$, rather than two
samples concentrated at $0.5$ and $0.8$ as originally planned. Because the
LCDCS detection algorithm was only sensitive to cluster light in the inner one
to two hundred kpc, our set of clusters has a broad range of richness and
internal structure on larger scales. As in all optically selected samples, our
cluster set contains a few superpositions of significant objects at differing
redshifts. The effect is smaller, however, than in earlier surveys based on
detecting light enhancements over larger areas (e.g. \citet{OPL98,dCetal99}).  The
velocity dispersions of our systems, as estimated both from our lensing maps
and from our spectroscopy, span the range from a few hundred to over one
thousand ${\rm km}\,{\rm s}^{-1}$. Our richest clusters are similar to those
detected by X-ray surveys for distant objects, but we also have less regular
and lower richness systems.  The diversity of the population is evident in the
isopleth maps and the colour-magnitude diagrams presented above. Not only do
the internal structure and the total galaxy number of the clusters vary
dramatically, but so also do the relative numbers of red and blue and of faint
and bright galaxies within them.  This diversity makes our sample ideal for
studying the environmental dependence of galaxy evolution, particularly when
our data at half the present age of the Universe are compared with the
low-redshift baseline now available with very large statistics and uniform
photometric and spectroscopic data from the Sloan Digital Sky Survey.  Such
studies of the evolution of the galaxy population form the basis for later
papers in this series.

\begin{acknowledgements}
We have benefitted from the generosity of the ESO/OPC.  We thank Henry
McCracken for providing us with catalogues from the CFDFs.
G.~D.~L. and S.~C. thank the Alexander von Humboldt Foundation, the Federal
Ministry of Education and Research, and the Programme for Investment
in the Future (ZIP) of the German Government for financial support.
G.~R. thanks Special Research Area No 375 of the German Research Foundation
for financial support.

\end{acknowledgements}
\bibliographystyle{aa}
\bibliography{white_ediscs}

\begin{figure*}
  \begin{center}
    \resizebox{16cm}{!}{\includegraphics{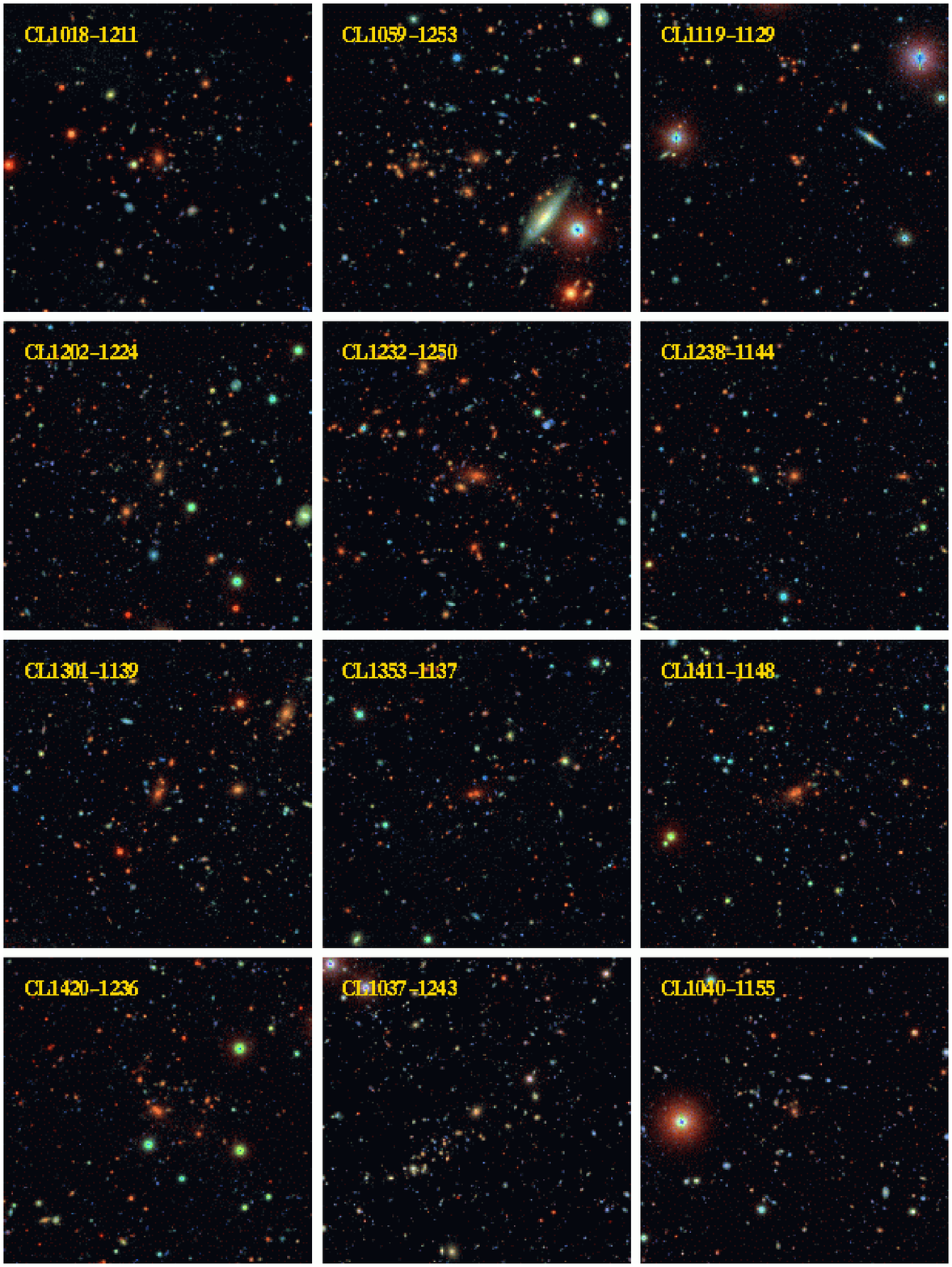}}
    \caption{}
    \label{fig:fig3}
  \end{center}
\end{figure*}
\addtocounter{figure}{-1}
\begin{figure*}
  \begin{center}
    \resizebox{\hsize}{!}{\includegraphics{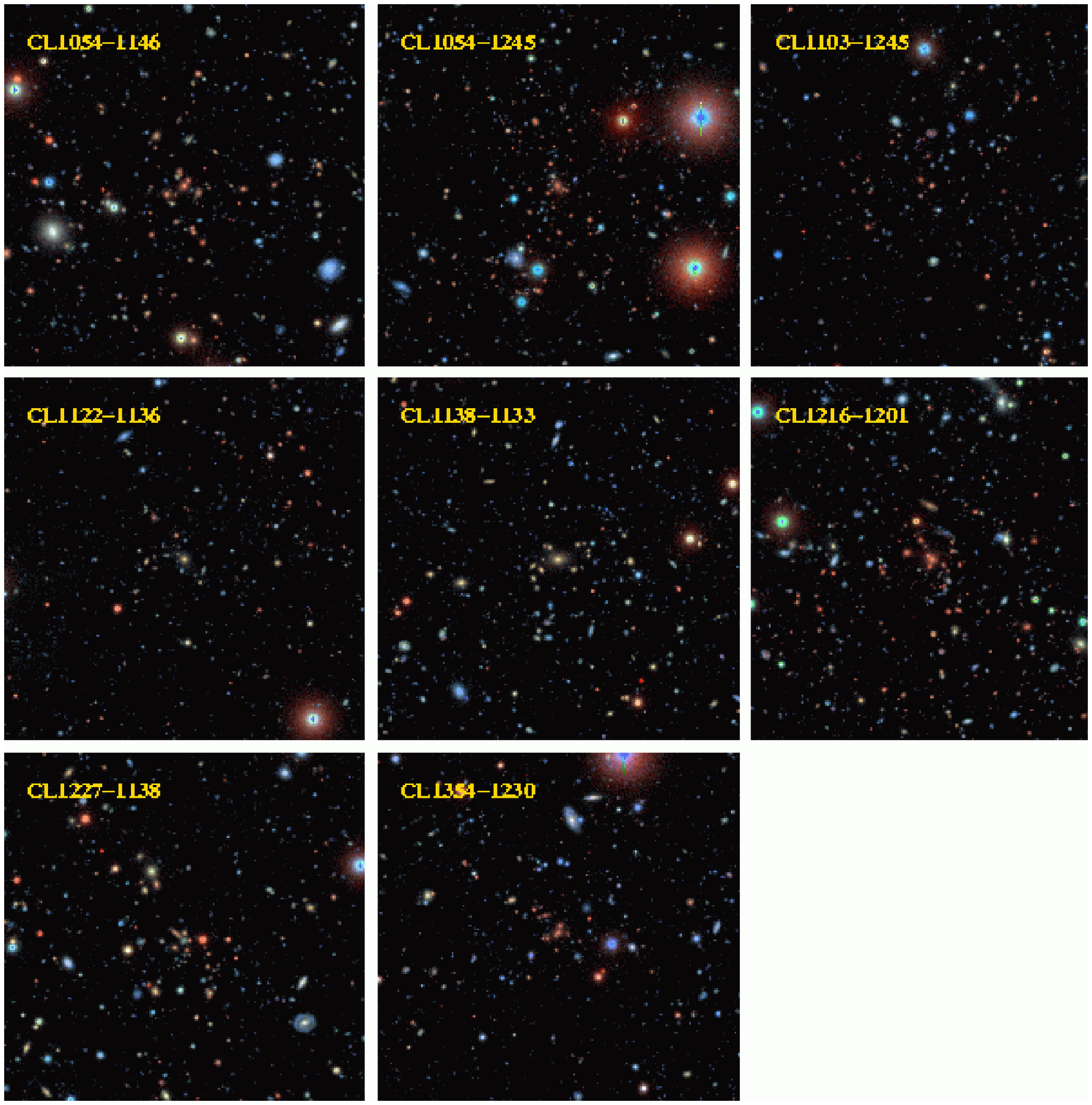}}
    \caption{--continued. False colour images of the central regions of each of
      our EDisCS fields. Each image is $3\arcmin\times 3\arcmin$ and is centred
      on our BCG candidate (except for cl1227.9$-$1138, see text). The images
      are made by combining the seeing matched images in our three optical
      bands with a colour stretch which maximizes the variation of image colour
      with galaxy spectral energy distribution.  These colour scales are the
      same within the mid-$z$ and within the hi-$z$ sample but differ between
      the two. In each image north is at the top and east is to the left.
      Images of our mid-$z$ sample in RA order are followed by images of our
      hi-$z$ sample also in RA order.}
  \end{center}
\end{figure*}

\begin{figure}
  \begin{center}
    \resizebox{\hsize}{!}{\includegraphics{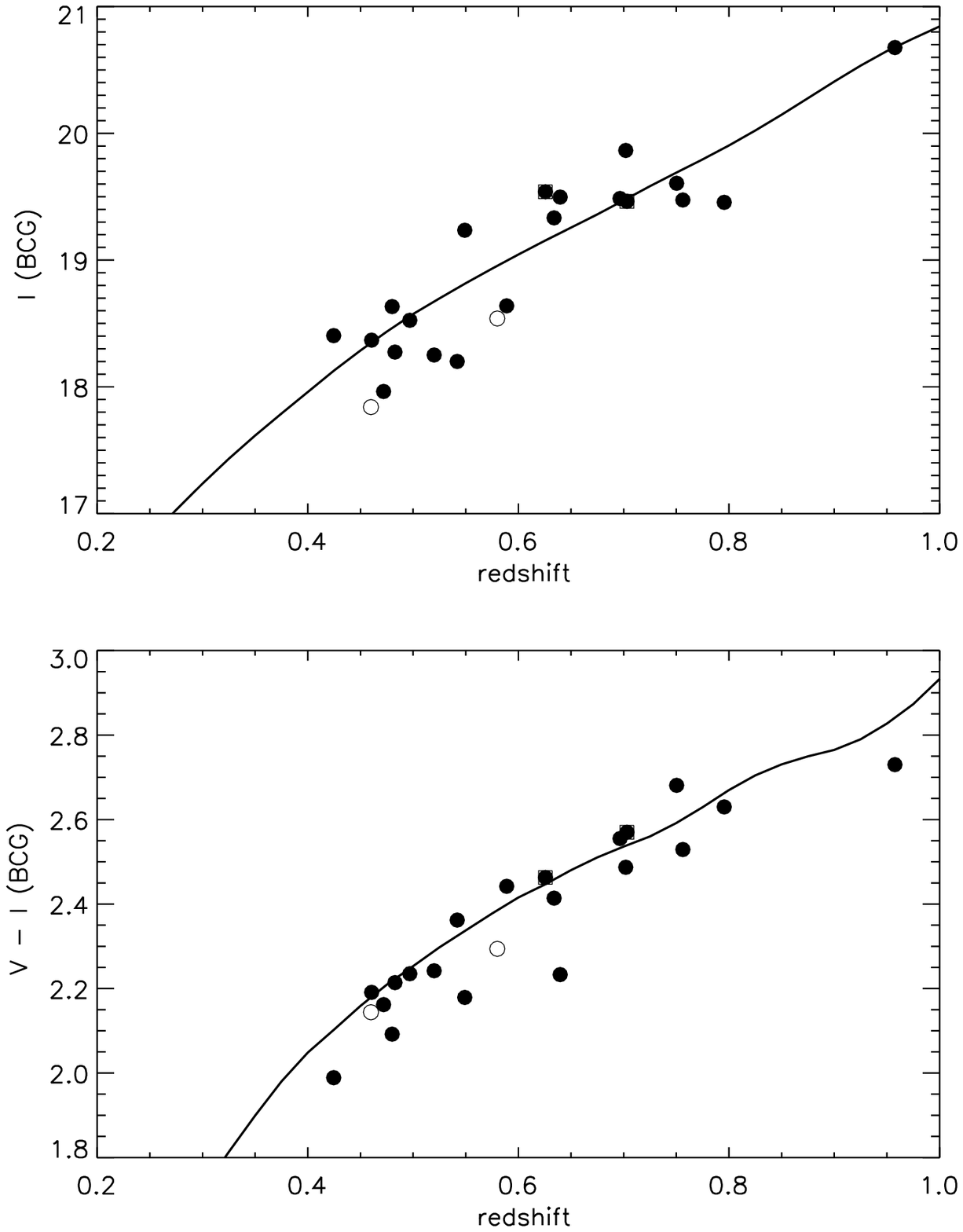}}\\%
    \caption{``Total'' I-band apparent magnitudes and ${\rm V}-{\rm I}$
      colours of our BCG candidates are plotted against redshift.  The lines
      in each plot are taken from a \citet{BC03} model which assumes passive
      evolution after a single, solar metallicity burst of star formation at
      $z=3$. The luminosity of the model is normalized so that at $z=0$ it
      reproduces the V-band absolute magnitude of NGC$4889$, the brightest
      galaxy in the Coma cluster.  In both plots, open symbols correspond to
      the clusters cl1059.2$-$1253 and cl1037.9$-$1243, for which we do not
      have a spectroscopic redshift for the object identified as BCG.  In
      these cases, we use the cluster redshift. The filled squares correspond
      to the BCG candidates in the two additional structures away from the
      LCDCS detection in field 1103.7$-$1245.}
    \label{fig:fig4}
  \end{center}
\end{figure}

\begin{figure*}
  \begin{center}
    \resizebox{\hsize}{!}{\includegraphics{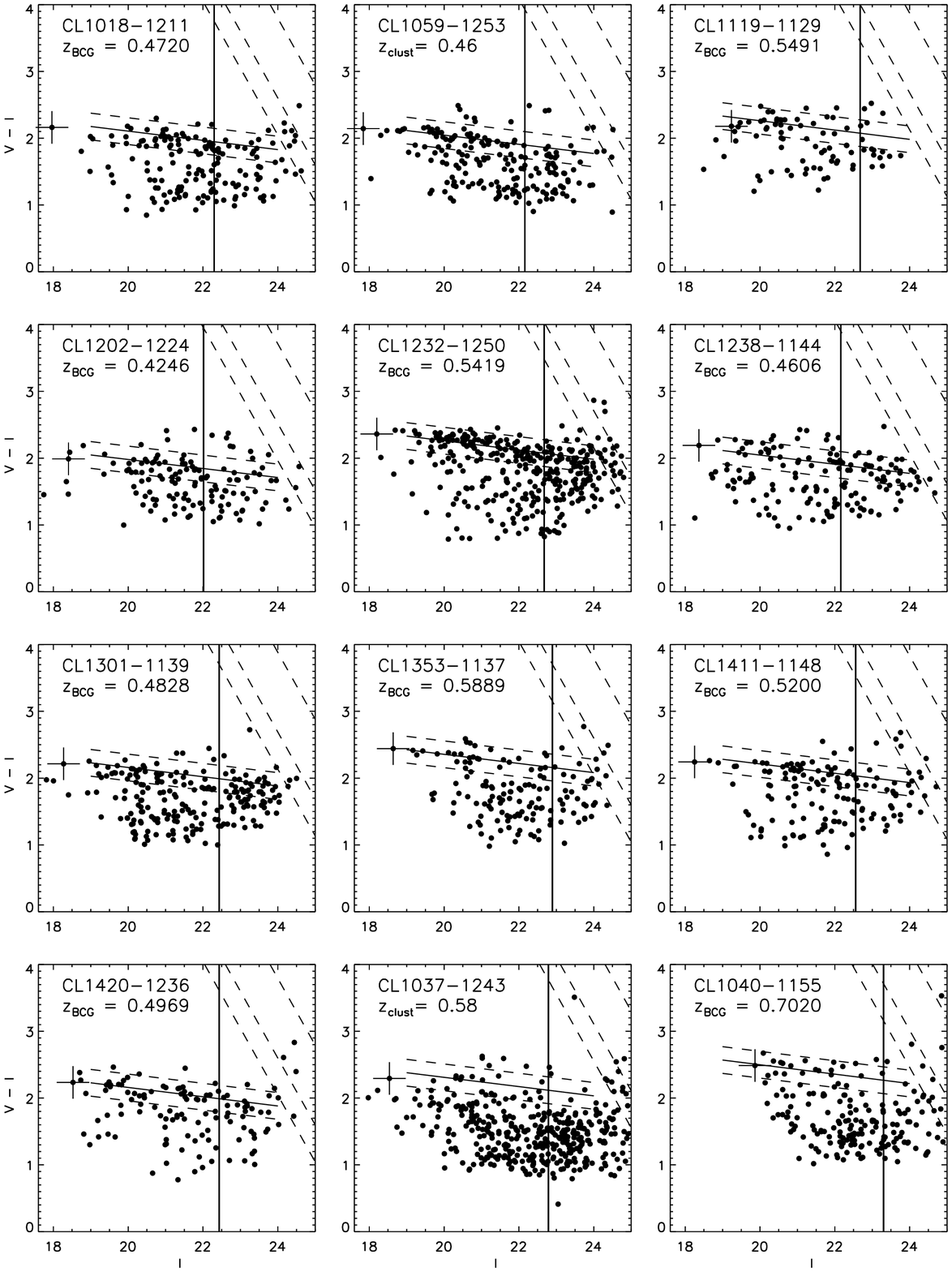}}
    \caption{}
    \label{fig:fig5}
  \end{center}
\end{figure*}
\addtocounter{figure}{-1}
\begin{figure*}
  \begin{center}
    \resizebox{\hsize}{!}{\includegraphics{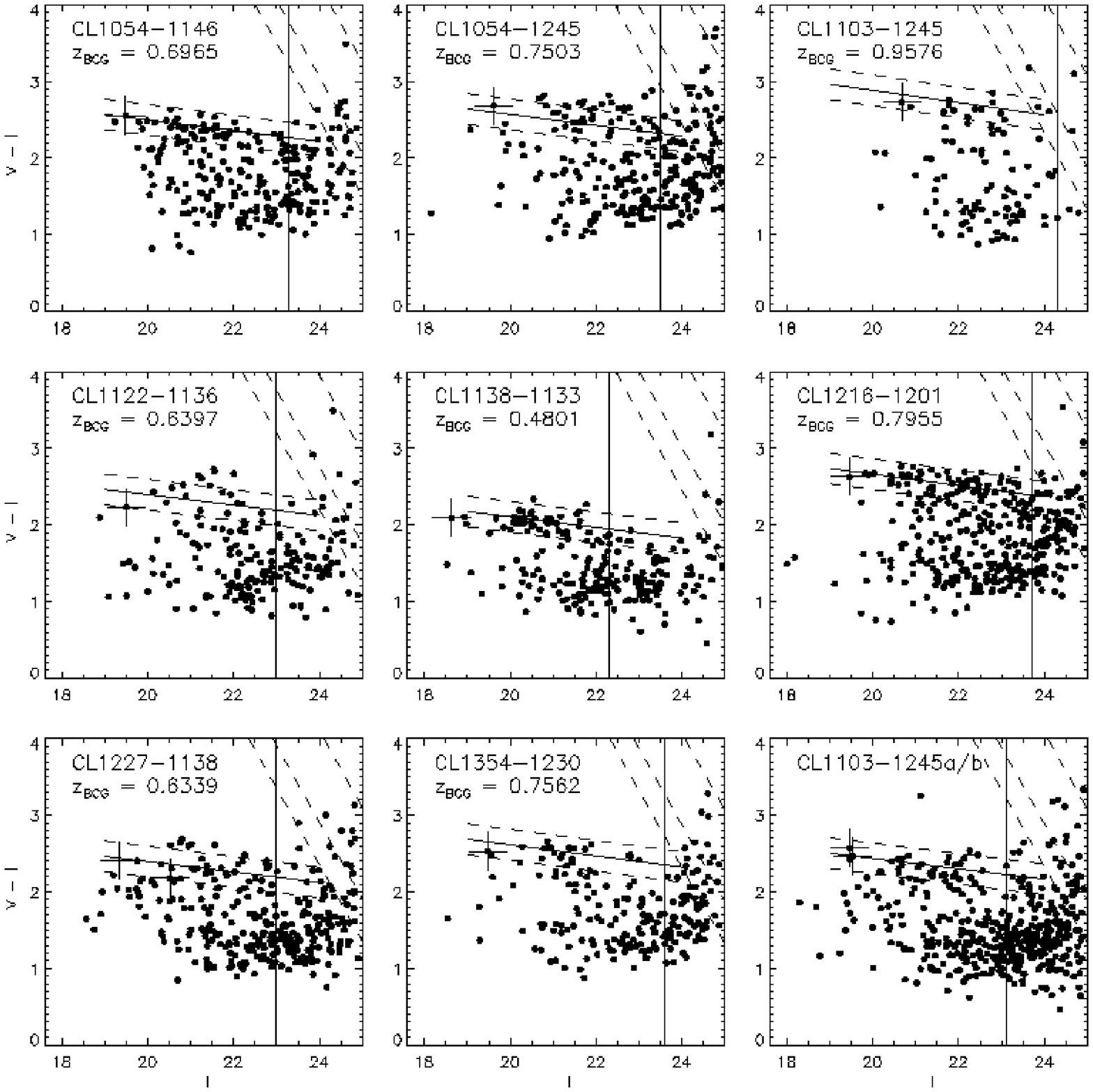}}    
    \caption{--continued.  ${\rm V}-{\rm I}$ versus I colour-magnitude diagrams
      for galaxies projected within $1.0$ Mpc of the BCG candidate (except for
      cl1227.9$-$1138, see text) in each of our $20$ EDisCS fields.  Galaxies
      for which our photometric redshift routines give a low probability of
      being near the cluster redshift have been eliminated. The BCG candidate
      itself is shown with a large cross. (Note that for cl1227.9$-$1138 the
      BCG candidate is outside our 1 Mpc circle.) The expected position of the
      red sequence is shown by a solid line, while a band within $\pm 0.2$
      magnitudes of this sequence is delineated by dashed lines. We use the
      count of galaxies within this band and brighter than the vertical solid
      line to estimate the richness of each cluster.  Steeply slanting dashed
      lines show $1$, $3$ and $5\sigma$ detection limits for our V-band
      photometry. The final panel shows a colour-magnitude plot for the two
      additional structures away from the LCDCS detection in the field
      cl1103.7$-$1245. Filled circles represent here all the galaxies within
      the largest square field for which we have both IR and optical data.}
  \end{center}
\end{figure*}

\begin{figure*}
  \begin{center}
    \resizebox{\hsize}{!}{\includegraphics{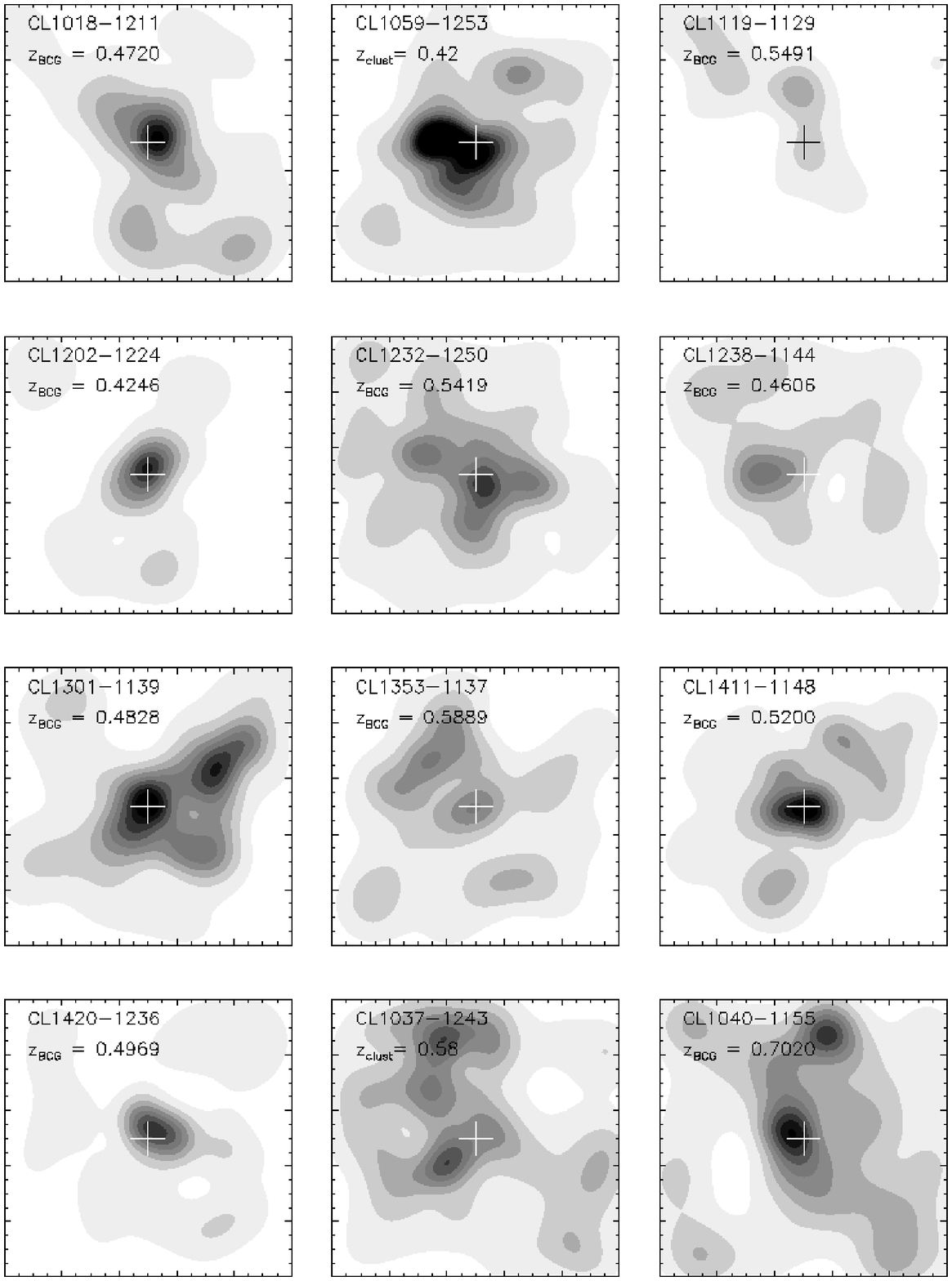}}
    \caption{}
    \label{fig:fig6}
  \end{center}
\end{figure*}
\addtocounter{figure}{-1}
\begin{figure*}
  \begin{center}
    \resizebox{\hsize}{!}{\includegraphics{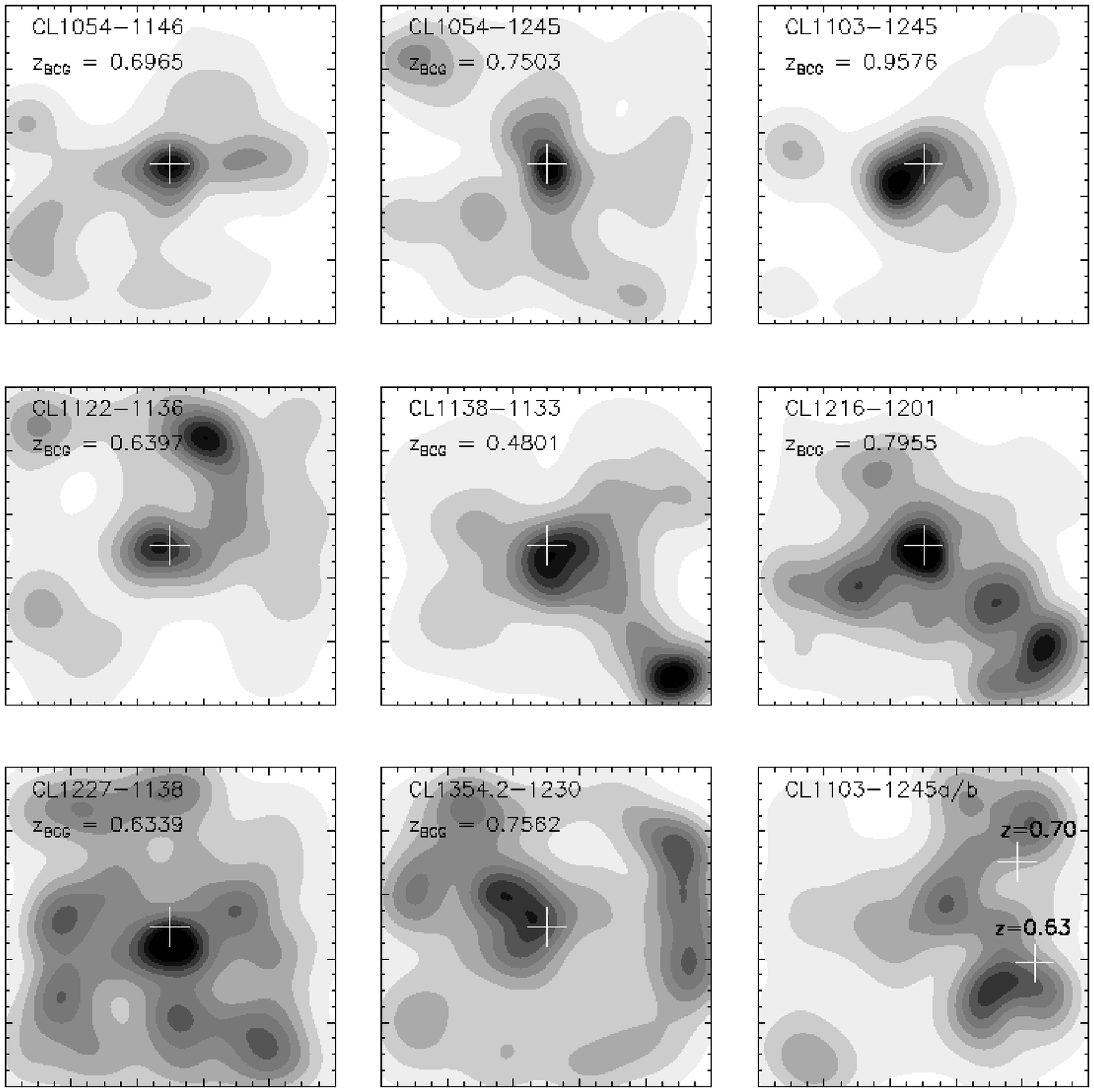}}
    \caption{-- continued.  Adaptively smoothed isopleths for the surface
      density of galaxies brighter than I $=25$ for a $4\farcm8\times 4\farcm8$
      EDisCS region centred on the BCG candidate (except for cl1227.9$-$1138,
      see text) in each of our fields. Stars and galaxies with photometric
      redshifts inconsistent with being close to the cluster redshift are
      eliminated before contouring. The lowest contour level plotted
      corresponds to $7.2$ galaxies per square arcminute.  Subsequent contours
      correspond to further steps of $7.2$ galaxies per square arcminute for
      clusters cl1232.5$-$1250, cl1037.9$-$1243, cl1054.4$-$1146,
      cl1054.7$-$1245 and cl1216.8$-$1201. In all other cases subsequent
      contours correspond to steps of half this size.  The position of our BCG
      candidate is noted in each plot by a cross. As in Fig.~\ref{fig:fig1},
      north is at the top and east at the left of each map. The final panel
      shows isopleths for the two additional structures away from the LCDCS
      detection in the field cl1103.7$-$1245.}
  \end{center}
\end{figure*}

\end{document}